\documentclass[preprint,showpacs,preprintnumbers]{revtex4}
\usepackage{amssymb}
\usepackage{amsmath}
\usepackage{graphicx}
\usepackage{dcolumn}
\usepackage{bm}
\usepackage{booktabs,array,tabularx,multirow}
\usepackage{epsfig}
\usepackage[T1]{fontenc}
\usepackage{ae,aecompl}
\usepackage{epstopdf, epsfig}
\usepackage{color}
\usepackage{appendix}
\usepackage[export]{adjustbox}
\usepackage{graphicx}
\usepackage{epstopdf, epsfig}
\usepackage[section]{placeins}
\usepackage[utf8]{inputenc}

\begin{document}

\title{Discrete Boltzmann modeling of Rayleigh-Taylor instability: Effects of interfacial tension, viscosity, and heat conductivity}
\author{Jie Chen$^{1,4}$, Aiguo Xu$^{1,2,3}$\footnote{
Corresponding author. E-mail: Xu\_Aiguo@iapcm.ac.cn}, Dawei Chen$^1$, Yudong Zhang$^5$, Zhihua Chen$^4$}
\affiliation{1, Laboratory of Computational Physics, Institute of Applied Physics and Computational Mathematics, P. O. Box 8009-26, Beijing 100088, China \\
2, State Key Laboratory of Explosion Science and Technology, Beijing Institute of Technology, Beijing 100081, China \\
3, HEDPS, Center for Applied Physics and Technology, and College of Engineering, Peking University, Beijing 100871, China \\
4, Key Laboratory of Transient Physics, Nanjing University of Science and Technology, Nanjing 210094, China \\
5, School of Mechanics and Safety Engineering, Zhengzhou University, Zhengzhou 450001, China }
\date{\today }

\begin{abstract}
The two-dimensional Rayleigh-Taylor Instability (RTI) in compressible flow with inter-molecular interactions is probed via the Discrete Boltzmann Method (DBM). The effects of interfacial tension, viscosity and heat conduction are investigated. It is found that the influences of interfacial tension on the
perturbation amplitude, bubble velocity, and two kinds of entropy production rates
all show differences at different stages of RTI evolution. It inhibits the RTI evolution at the bubble acceleration stage, while at the asymptotic velocity stage, it first promotes and then inhibits the RTI evolution.
Viscosity and heat conduction inhibit the RTI evolution. Viscosity shows a suppressive effect on the entropy generation rate related to heat flow at the early stage but a first promotive and then suppressive effect on the entropy generation rate related to heat flow at a later stage. Heat conduction shows a promotive effect on the entropy generation rate related to heat flow at an early stage. Still, it offers a first promotive and then suppressive effect on the entropy generation rate related to heat flow at a later stage. By introducing the morphological boundary length, we find that the stage of exponential growth of the interface length with time corresponds to the bubble acceleration stage. The first maximum point of the interface length change rate and the first maximum point of the change rate of the entropy generation rate related to viscous stress can be used as a criterion for RTI to enter the asymptotic velocity stage.

\pacs{47.11.-j, 47.20.-k, 05.20.Dd}
\end{abstract}

\maketitle

\preprint{APS/123-QED}

\section{Introduction}

Hydrodynamic instabilities are found in a wide range of natural sciences and engineering, such as solar helium, space ionospheric flows, inertial confinement fusion (ICF), pulsed burst engines and super-combustion ram engines, and even in artistic creation\cite{ZHOU20171,zhou2017rayleigh,Zhou2019Turbulent,zhou2021rayleigh}. Rayleigh-Taylor Instability(RTI), Richtmyer-Meshkov instability (RMI), and Kelvin-Helmhotz instability are all common hydrodynamic instabilities. RTI refers to the instability when a light medium supports a heavy medium or when a light medium accelerates a heavy medium\cite{rayleigh1882investigation,taylor1950instability}. RMI refers to the instability when a shock wave passes through the material interface\cite{richtmyer1960taylor,meshkov1969instability}. RMI can also be considered a special case of RTI that can be treated as an impulsive analog of the RTI. The presence of hydrodynamic instabilities can be beneficial in some cases, such as aero-engine combustion, where the presence of hydrodynamic instabilities facilitates fuel mixing and improves the combustion efficiency of the engine. It can also be undesirable in other cases, such as inertial confinement fusion, where hydrodynamic instabilities and the non-linear complex flows they cause can seriously affect ICF ignition and cause fusion performance cliff problems\cite{lifeng2021review}. Whether from an interest in natural science or the need for engineering applications, experts and scholars from all backgrounds are working on problems that interest them in this highly multidisciplinary field.

As the research progresses, the problem being studied and the models being used are becoming more and more practical and complex\cite{luo2019effects,zou2019richtmyer,liang2019direct,ge2020late,Yao2020Kinetic,Manuel2021On,Cai2021Hybrid,wu2021refined,liang2021late,
liang2021universal,vadivukkarasan2021temporal,li2021growth,ding2021convergent,liang2022phase,
wang2022transition,zhou2022terminal,Li2022Instability}. In recent years, the results in statistical physics perspectives obtained with the help of the Discrete Boltzmann Method (DBM) and complex physical field analysis techniques have also provided a range of new insights into the study of hydrodynamic instabilities. Examples include the relationship between the nonuniformity of various macroscopic quantities and different forms of nonequilibrium behavior in terms of correlations\cite{chen2016viscosity,chen2018collaboration}, the entropy of mixing during RTI development, and the relationship between the mixed entropy and nonequilibrium characteristic quantities of a system\cite{lin2017discrete} and employing non-equilibrium characteristic quantities and morphological characteristics to trace and characterize the instabilities interfaces, the mixed layer widths, the stages of development and the rates of growth\cite{gan2019nonequilibrium,zhang2019discrete,chen2020morphological}.

The actual medium has an intermolecular interaction potential. Suppose the intermolecular interaction potential is modest compared to the external force acting on the medium. In that case, we can temporarily overlook the effect of the intermolecular interaction potential and capture the problem's primary paradox, which can also help us get a relatively satisfactory answer. However, when the intermolecular interaction potential is no longer negligible, the case is more complex. It will lead to complex physical processes such as perturbation stabilization, tensile fracture, and melting phase transitions\cite{miles1966taylor,piriz2009linear,jian2017stability,bi2020experimental}. As another example, in creating jet models and theories for microjet problems, which are significant in engineering domains such as hypervelocity impact, ICF ignition and implosion, it is also crucial to consider the strength effects of materials\cite{anmin2018theoretical}. It is known that in many RTI studies intermolecular interactions were neglected and the ideal gas equation of state was used. However, the intermolecular interaction potential in a fluid medium can also cause some unique events in the evolution of interfacial instability. The difference in the intermolecular interaction potential between two fluids causes the well-known interfacial tension. It is a manifestation of the medium strength effect at the medium interface.

The existence of interfacial tension will also result in a cutoff wavelength for the development of RTI. The perturbation will be stabilized when the wavelength is smaller than the cut-off wavelength\cite{mikaelian1990rayleigh}. The study of Banerjee and Kanjilal also reflects this fact from another perspective\cite{banerjee2015effect}. They found that when the interfacial tension was less than the critical value, the degree of inhibition of the bubble tip growth rate was highly dependent on the interfacial tension. On the other hand, oscillatory behavior of the interface was observed when the interfacial tension was greater than the critical threshold. The Atwood number and the vortex volume's intensity are both related to the critical value of interfacial tension. Both theoretical analysis and numerical simulations show that interfacial tension inhibits bubble growth during the linear stage of RTI\cite{xin2013bubble,yuan2014theoretical,Ningning2016NumericalSimulation}. During the non-linear stage of RTI development, most scholars' studies showed that interfacial tension decreases the asymptotic velocity of bubbles\cite{yuan2014theoretical,sohn2009effects,li2016effect,tong2013effects}, however, Huang et al. showed that the growth of bubble amplitude does not always slow down with increasing interfacial tension but is promoted within a small range of interfacial tension\cite{hao2021effect}.
In addition to its effect on bubble growth rate and amplitude, interfacial tension also exhibits effects such as squeezing of spikes, inhibition of interfacial coiling, and induction of interfacial pinch-off to form discrete droplets\cite{daly1969numerical,zhang2000interface,matsuoka2009vortex}. Capillary waves, squeezing, and rupture during the evolution of RTI are also caused by interfacial tension\cite{shin2022numerical}. It has been suggested that the formation of some heavy fluid droplets, which fall freely into the vacuum in the early stages of RTI, is also associated with interfacial tension cutting continuous surfaces\cite{zufiria1988bubble}. For the turbulent mixing of RTI, Young and Ham's results show that interfacial tension reduces the effective mixing rate of multiple bubbles, decreases the flow's anisotropy, and induces more homogeneous mixing\cite{young2006surface}. Sohn and Beak found that the mixing rate of bubbles decreases with increasing interfacial tension at the early nonlinear stage and increases with increasing interfacial tension at the self-similar stage\cite{sohn2017bubble}. Previous studies mainly focused on the effect of interfacial tension on the growth rate and amplitude of RTI development, the turbulent mixing rates, etc. Furthermore, these studies are based on various assumptions such as inviscid, incompressible, isothermal, etc.
Current hydrodynamic instabilities researches using DBM are still in the early stages. They are all based on the relatively simple ideal gas model which ignores the influence of intermolecular interaction potential. It should provide new insights into statistical physics and complex physical field analysis perspectives and valuable additions to the existing research findings when introducing the intermolecular interaction potential in the DBM to study the hydrodynamic instabilities.

In this paper, we use the DBM that considers the intermolecular interaction potential and combine it with morphological analysis to provide insight into the evolution of two-dimensional RTI from a statistical physics perspective. In Section II, we describe the DBM modeling approach used in this paper. In Section III, we change the interfacial tension coefficient, viscosity coefficient, and heat conduction coefficient to study the effects of interfacial tension, viscosity, and heat conduction on the evolution of RTI. Section IV provides conclusions and a summaries.

\section{NUMERCAL METHODS}

\subsection{Discrete Boltzmann Method}

The discrete Boltzmann method is a mesoscale kinetic modeling method that has been rapidly developed in recent years. Unlike traditional fluid modeling methods, it is not based on the continuous medium assumption and near-equilibrium approximation. It can describe more rationally the local non-continuity and discrete effects due to small structures and local Thermodynamic Non-Equilibrium (TNE) effects due to fast flow modes. 
Since the description of system behavior needs more physical variables with increasing the non-continuity and/or TNE effects, in 2012, Xu, et al. \cite{xu2012lattice} proposed to use the non-conservative moments of $\left( f_{i}-{f^{\text{eq}}_{i}} \right)$ to
describe how and how much the system deviates from the thermodynamic equilibrium state, and to check the corresponding TNE effects, where $f_i$ is the discrete distribution function and $f^{\text{eq}}_i$ is the corresponding equilibrium, $i$ is the index of discrete velocity.
In 2015, they \cite{xu2015multiple} proposed to use the non-conservative moments of $\left( f_{i}-{f^{\text{eq}}_{i}} \right)$ to open phase space. The phase space and subspaces are used to describe the TNE states and behaviors. The distance between a state point to the origin is used to describe the TNE strength from a perspective. In 2018 \cite{Xu2018-RGD31}, the distance $D$ between two state points were used to describe the difference of two states deviating from their thermodynamic equilibriums. The reciprocal of the distance, $S = 1/D$, is defined as the similarity of the two states deviating from their thermodynamic equilibriums. The mean value of $D$ within a time interval, $\bar{D}$, is used to roughly describe the difference of the two kinetic processes from a perspective, and the reciprocal of the mean distance, $S^{P} = 1/\bar{D}$, is defined as the similarity of the two kinetic processes. 
Therefore, DBM is a further development of the statistical physical phase space description method in the form of the discrete Boltzmann equation.
In 2021\cite{xu2021modeling}, the phase-space description method was further extended to any system characteristics. A set of (independent) characteristic quantities is used to open phase space, and this space and its subspaces are used to describe the system properties.  A point in the phase space corresponds to a set of characteristic behaviors of the system. Distance concepts in the phase space or its sub-spaces are used to describe the difference and similarity of behaviors.
The DBM makes it clear that any definition of non-equilibrium strength is dependent on the perspective of study, and that the non-equilibrium behavior of complex flows needs to be studied from multiple perspectives. The DBM can go beyond traditional fluid modeling in terms of both depth and breadth of the description of non-equilibrium behavior\cite{xu2021progress,xu2021Progressofmesoscale,xu2021modeling,gan2022multiphase}.

The DBM model is built in a three-step process. The first step is to introduce an appropriate kinetic equation with simple collision operator. The second step is the discretization of the velocity space. The third step is the description of nonequilibrium states and nonequilibrium behaviour. Among them, the first two steps are coarse-grained physical modeling, which requires that the system behaviours concerned remain during the process of model simplification. The third step is the purpose and core of the DBM. Based on the fact that (i) the intermolecular interaction potential is far from weak and simple as required by the Boltzmann equation and (ii) the degree of non-equilibrium concerned may be far beyond the quasi-equilibrium required by the original Bhatnagar-Gross-Krook (BGK) model\cite{bhatnagar1954model}, the kinetic behaviour of most systems cannot be studied by the original BGK kinetic theory alone. Consequently, the BGK-like model used in practice can be regarded as a modified version based on mean-field theory.  The main responsibilities of the mean-field theory are twofold: (i) to supplement the description of the intermolecular interaction potential effect that the Boltzmann equation misses and (ii) to modify the application range of the BGK-like model so that it can be extended to a higher degree of nonequilibrium\cite{gan2022multiphase}.

The DBM has been successfully applied to the study of fluid instability problems, and a series of research results and insights have been gained with its help \cite{lai2016nonequilibrium,de2018discrete,zhang2021delineation,chen2021specific,chen2022effects}.
Including the Van Der Waals (VDW) equation of state in the existing discrete Boltzmann BGK model for RTI systems and considering the contribution of the density gradient to the entropy and internal energy of the system, we can obtain the DBM model as
\begin{equation}
{{\partial }_{t}}{{f}_{ji}}+{{\mathbf{v}}_{ji}}\cdot \nabla {{f}_{ji}}-\frac{\mathbf{a}\cdot \left( {{\mathbf{v}}_{ji}}-\mathbf{u} \right)}{RT}f_{ji}^{\text{eq}}=-\frac{1}{\tau }\left[ {{f}_{ji}}-f_{ji}^{\text{eq}} \right]+{{I}_{ji}}\text{,}  \label{Eq.1}
\end{equation}%
where ${{f}_{ji}}$ and $f_{ji}^{\text{eq}} $ are the discrete distribution function and the discrete equilibrium distribution function, respectively, and ${{I}_{ji}} $ is the intermolecular interaction potential term\cite{gonnella2007lattice,ZhangYudong2019Modelingandresearchofunbalanced},
\begin{equation}
{{I}_{ji}}=-\left[ A+\mathbf{B}\cdot\left( {{\mathbf{v}}_{ji}}-\mathbf{u} \right)+\left( C+{{C}_{q}} \right)\left( {{\mathbf{v}}_{ji}}-\mathbf{u} \right)\cdot \left( {{\mathbf{v}}_{ji}}-\mathbf{u} \right) \right]f_{ji}^{\text{eq}}\text{,}  \label{Eq.2}
\end{equation}%
where
\begin{equation}
A=-2\left( C\text{+}{{C}_{q}} \right)T\text{,}  \label{Eq.3}
\end{equation}%
\begin{equation}
\mathbf{B}=\frac{1}{\rho T}\nabla \cdot \left[ \left( P-\rho T \right)\mathbf{I}+\bm{\Lambda } \right]\text{,}  \label{Eq.4}
\end{equation}%
\begin{equation}
\begin{split}
 \text{ } C=\frac{1}{2\rho {{T}^{2}}}\left[ \left( P-\rho T \right)\nabla \cdot \mathbf{u}+\bm{\Lambda }:\nabla \mathbf{u}+a{{\rho }^{2}}\nabla \cdot \mathbf{u} \right. \\
\text{ } \left. -K(\frac{1}{2}\nabla \rho \cdot \nabla \rho \nabla \cdot \mathbf{u}+\rho \nabla \rho \cdot \nabla \left( \nabla \cdot \mathbf{u} \right)+\nabla \rho \cdot \nabla \mathbf{u}\cdot \nabla \rho ) \right]\text{,} \\
\end{split} \label{Eq.5}
\end{equation}%
\begin{equation}
{{C}_{q}}=\frac{1}{\rho {{T}^{2}}}\nabla \cdot \left( q\rho T\nabla T \right)\text{,}  \label{Eq.6}
\end{equation}%
where
\begin{equation}
\bm{\Lambda }={{p}_{1}}\mathbf{I}+M\nabla n\nabla n\text{,}  \label{Eq.7}
\end{equation}%
\begin{equation}
\begin{split}
\begin{aligned}
\text{ } {{p}_{1}}&=-nT\nabla \frac{M}{T}\cdot \nabla n-nM{{\nabla }^{2}}n+(n{M}'-M)\frac{1}{2}{{\left| \nabla n \right|}^{2}} \\
&=n{{\mu }_{b}}-{{e}_{b}}+T{{s}_{b}}-P \text{,}\\
 \end{aligned}
\end{split}
 \label{Eq.8}
\end{equation}%
\begin{equation}
P=\frac{\rho T}{\text{1}-b\rho }-a{{\rho }^{2}}\text{.}  \label{Eq.9}
\end{equation}%

In the DBM, we can describe the non-equilibrium behaviour and characteristics of complex flows with the help of non-conservative moments of $\left( {{f}_{ji}}-f_{ji}^{\text{eq}} \right)$,
\begin{equation}
\bm{\Delta }_{n}^{*}=\mathbf{M}_{n}^{*}\left( {{f}_{ji}} \right)-\mathbf{M}_{n}^{*}\left( f_{ji}^{\text{eq}} \right)\text{,}  \label{Eq.10}
\end{equation}%
where $\mathbf{M}_{n}^{*}\left( {{f}_{ji}} \right)$ and $\mathbf{M}_{n}^{*}\left( f_{ji}^{\text{eq}} \right)$ are the $nth$ order kinetic moments of the discrete distribution functions ${{f}_{ji}}$ and $f_{ji}^{\text{eq}}$ with respect to the molecular rise and fall velocity $\left( {{\mathbf{v}}_{ji}}-\mathbf{u} \right)$, respectively, i.e. the $nth$ order central moment of ${{f}_{ji}}$ and $f_{ji}^{\text{eq}}$. $\mathbf{\Delta }_{n}^{*}$ is called the TNE characteristic quantity of the system. TNE quantities of different orders describe the non-equilibrium state of the system from their own respect. For example, ${\Delta }_{2\alpha \beta }^{*}$ denotes viscous stress and ${\Delta }_{\left( 3,1 \right)\alpha }^{*}$ denotes heat flow. The subscripts ``2" and ``$\left( 3,1 \right)$" denote the $2\text{nd}$ order tensor and the $1\text{st}$ order tensor reduced from the $3\text{rd}$ order tensor, respectively. The subscripts ``$\alpha \beta $" and ``$\alpha $" denote the coordinate components of the tensor.

With the help of these non-equilibrium quantities, we can study the entropy increase of the system, the spatial correlation of the system, the temporal correlation of the system, the spatio-temporal correlation of the system, and the competition and cooperation between different non-equilibrium behaviors of the system\cite{zhang2019entropy}. Using the phase space and its sub-spaces opened by the non-equilibrium characteristic quantity, we can study the degree or intensity of state deviation from equilibrium, the similarity of two non-equilibrium states, the similarity of two kinetic processes, etc.\cite{2015Progess,Xu2018-Chap2,xu2015multiple,xu2016complex}

\subsection{Entropy production of system}

When considering the contribution of the density gradient to entropy and internal energy, the generalized form of the total entropy of the system can be written as\cite{gonnella2007lattice,onuki2005dynamic}
\begin{equation}
{{S}_{b}}=\int{\left[ ns\left( n,e \right)-\frac{1}{2}C{{\left| \nabla n \right|}^{2}} \right]}d\mathbf{r}\text{,}  \label{Eq.11}
\end{equation}%
where the space integrals extend to the entire computational domain, $n$ is the particle number density, $s$ is the entropy per particle, $C$ is a constant and the gradient term represents a decrease of entropy because of the inhomogeneity of $n$.
Then, we can get
\begin{equation}
\frac{d{{S}_{b}}}{dt}=\int{\left( \frac{1}{T}\nabla \cdot \mathbf{j}+\frac{1}{T}\bm{\Pi }:\nabla \mathbf{u} \right)}d\mathbf{r}\text{,}  \label{Eq.12}
\end{equation}%
where $\mathbf{j}$ and $\bm{\Pi }$ are heat flux and viscous stress, respectively.
The time partial derivative of the entropy density can be obtained from the relationship of the material derivatives,
\begin{equation}
\begin{split}
\frac{\partial {{s}_{b}}}{\partial t} =-\nabla \cdot \left( \mathbf{u}{{s}_{b}}-\frac{1}{T}\mathbf{j} \right)-\mathbf{j}\cdot \nabla (\frac{1}{T})+\frac{1}{T}\bm{\Pi }:\nabla \mathbf{u} \\
\end{split}
\text{.}  \label{Eq.13}
\end{equation}%

The entropy production rate can be divided into two parts. One part is contributed by the heat flow
\begin{equation}
{{\dot{S}}_{\text{NOEF}}}=\int{\left( \mathbf{\Delta }_{\left( 3,1 \right)}^{*}-{{c}_{p}}q\rho T \right)\cdot \nabla \frac{1}{T}}d\mathbf{r}  \label{Eq.14}
\end{equation}%
and the other part is contributed by the viscous stress
\begin{equation}
{{\dot{S}}_{\text{NOMF}}}=\int{-\frac{1}{T}\mathbf{\Delta }_{2}^{*}:\nabla \mathbf{u}}d\mathbf{r}\text{.}  \label{Eq.15}
\end{equation}%
In the continuous limit, $\mathbf{\Delta }_{\left( 3,1 \right)}^{*}-{{c}_{p}}q\rho T\text{=}-\kappa\nabla T$, where ${{c}_{p}}$ is the isobaric specific heat capacity, $\kappa$ is the heat conductivity, and $\mathbf{\Delta }_{2}^{*}=-\mu \left[ \nabla \mathbf{u}+{{\left( \nabla \mathbf{u} \right)}^{T}}-\left( \nabla \cdot \mathbf{u} \right)\mathbf{I} \right]$, where $\mu$ is the viscosity coefficient. Substituting them into Eqs. (\ref{Eq.14}) and (\ref{Eq.15}) respectively, we get that
\begin{equation}
\begin{split}
 \text{ } {{{\dot{S}}}_{\text{NOEF}}}=\int{{\kappa{{\left| \nabla T \right|}^{2}}}/{{{T}^{2}}}\;}d\mathbf{r} \\
\end{split}
\text{,}  \label{Eq.16}
\end{equation}%
\begin{equation}
\begin{split}
{{{\dot{S}}}_{\text{NOMF}}}=\int{\mu \left[ { \nabla \mathbf{u}:\nabla \mathbf{u}}+{{\left( \nabla \mathbf{u} \right)}^{T}}:\nabla \mathbf{u}-{{\left| \nabla \cdot \mathbf{u} \right|}^{2}} \right]/T\ }d\mathbf{r} \\
\end{split}
\text{.}  \label{Eq.17}
\end{equation}%

\section{\protect\bigskip Results and discussion}
We investigated the two-dimensional RTI flow by the DBM that considers the interfacial tension. In a computational domain with a width of $0.\text{256}$ and a height of $\text{1}\text{.024}$, the upper half is heavy fluid and the lower half is lighter fluid. At the initial time, an initial disturbance ${{y}_{c}}\left( x \right)={{y}_{0}}\cos \left( kx \right)$ of amplitude ${{y}_{0}}=0.05d$ exists at the light-heavy fluid interface that is at y=0, where $d$ is the width of the computational domain, $k=2\pi /\lambda $ is the wave number and $\lambda =d$ is the wavelength of the disturbance. The computational domain grid size is $256\times 1024$, and the grid length is $0.001$  in the $x$ and $y$ directions. The time step is $\text{6}\times \text{1}{{\text{0}}^{-\text{6}}}$ . The upper and lower boundaries are set as wall boundary conditions, and the left and right boundaries are set as periodic boundary conditions. The temperature of the light and heavy fluids at the initial moment are set to be constant of
\begin{equation}
\begin{aligned}
 \text{ } {{T}_{0}}\left( y \right)&=1.5,y>{{y}_{c}}\left( x \right) \\
 \text{ } {{T}_{0}}\left( y \right)&=3.0,y\le{{y}_{c}}\left( x \right)\text{.} \\
\end{aligned}
  \label{Eq.18}
\end{equation}%
The density at the top of the flow field is set to be 0.667, i.e.
\begin{equation}
{{\rho }_{0}}\left( 1.024 \right)=0.667\text{.}  \label{Eq.19}
\end{equation}%
Fluid in the flow field satisfies hydrostatic equilibrium, i.e.
\begin{equation}
{{\partial }_{y}}{{P}_{0}}\left( y \right)=-g{{\rho }_{0}}\left( y \right)\text{.}  \label{Eq.20}
\end{equation}%
Substituting the VDW equation of state into Eq.(\ref{Eq.20}) gives
\begin{equation}
{{\partial }_{y}}{{\rho }_{0}}\left( y \right)=g/2a\ -\frac{1}{2a}\frac{Tg}{T-2a{{\rho }_{0}}\left( y \right){{\left( 1-b{{\rho }_{0}}\left( y \right) \right)}^{2}}}\text{,}  \label{Eq.21}
\end{equation}%
where $a={9}/{8}\;,b={1}/{3}\;$ in this paper.

\subsection{Effect of interfacial tension on RTI}
We first investigated the effect of interfacial tension on the evolution of RTI. In this section, the gravitational acceleration is $g=0.2$, the relaxation time is $\tau =4\times \text{1}{{\text{0}}^{-5}}$, and the Prandtl number is $\rm\Pr =0.6$. The cases and parameter settings can be found in Table \ref{table1} in the Appendix C. Figure \ref{Fig1} shows the density contour of the flow field at different times for an interfacial tension coefficient of $K =1\times \text{1}{{\text{0}}^{-5}}$.

\begin{figure}[tbp]
\includegraphics[width=0.90\textwidth,trim=0.1 0.1 0.1 0.1,clip]{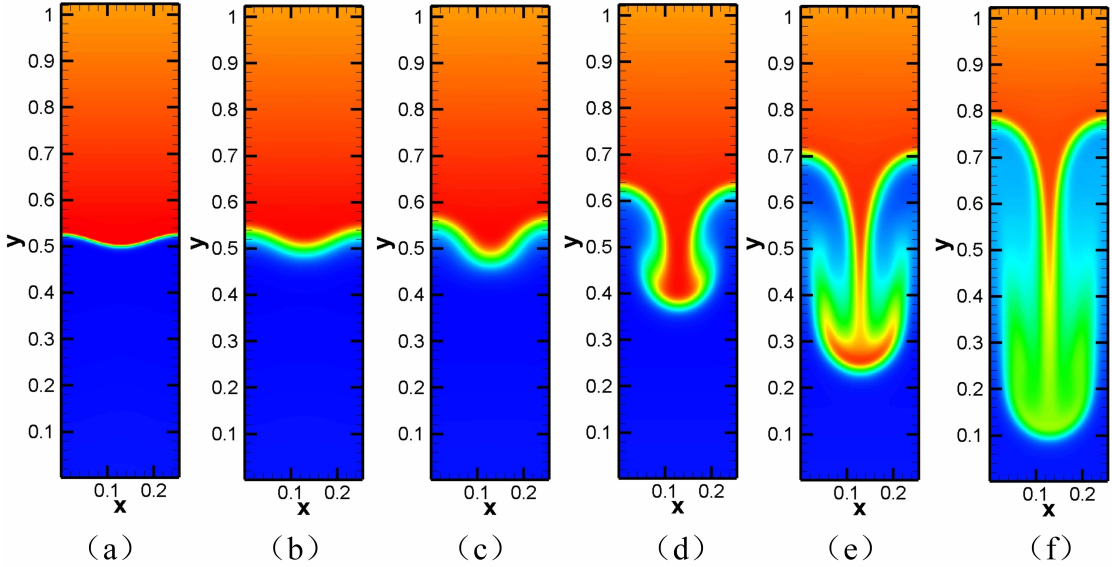}
\caption{Density contour of the flow field at different times for an interfacial tension coefficient of $K =1\times \text{1}{{\text{0}}^{-5}}$: (a) $t=0$, (b) $t=1.008$, (c) $t=2.016$, (d) $t=3.948$, (e) $t=6$, (f) $t=8.016$. Densities range within $\left[ 0.29,0.73 \right]$, increasing from blue to red.} \label{Fig1}
\end{figure}

\begin{figure}[tbp]
\center{\epsfig{file=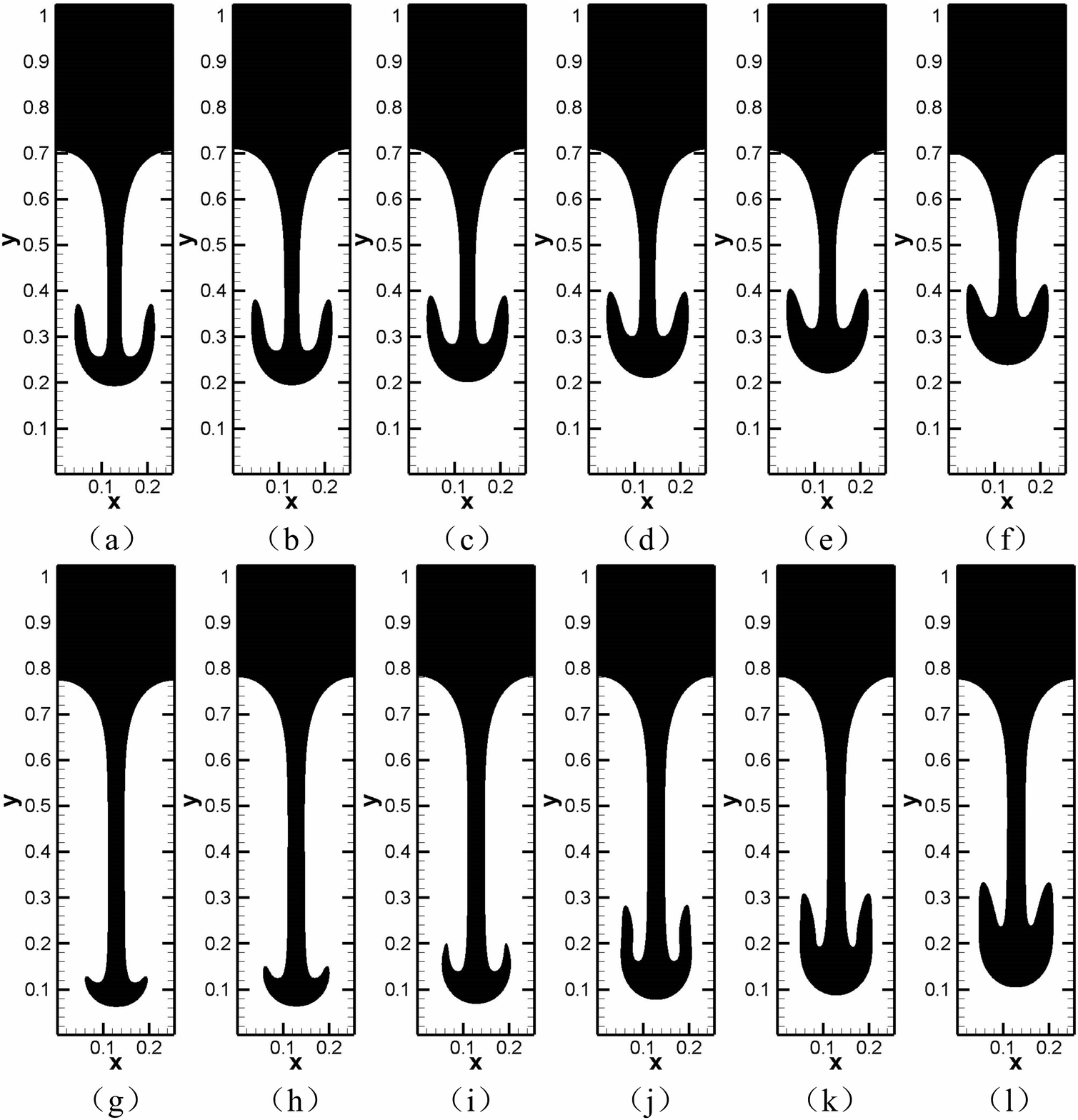,
width=0.9\textwidth }}
\caption{Evolution of RTI for different interfacial tension coefficients at $K =0$, $2\times \text{1}{{\text{0}}^{-6}}$, $4\times \text{1}{{\text{0}}^{-6}}$, $6\times \text{1}{{\text{0}}^{-6}}$, $8\times \text{1}{{\text{0}}^{-6}}$, and $1\times \text{1}{{\text{0}}^{-5}}$ for (a)-(f) $t=6$ and (g)-(l) for $t=8.016$.} \label{Fig2}
\end{figure}

Figure \ref{Fig2} shows the flow field at $t=6$ and $8.016$ for interfacial tension coefficients $K =0$, $2\times \text{1}{{\text{0}}^{-6}}$, $4\times \text{1}{{\text{0}}^{-6}}$, $6\times \text{1}{{\text{0}}^{-6}}$, $8\times \text{1}{{\text{0}}^{-6}}$, and $1\times \text{1}{{\text{0}}^{-5}}$. The average value of the light fluid density and the heavy fluid density at the initial interface is used to capture the interface of the light fluid and the heavy fluid. RTI evolution becomes increasingly delayed, and the ``mushroom" structure becomes more ``full'' with the increasing interfacial tension coefficient, as illustrated in Fig. \ref{Fig2}.

\begin{figure}[tbp]
\includegraphics[width=0.8\textwidth,trim=0 0 0 0,clip]{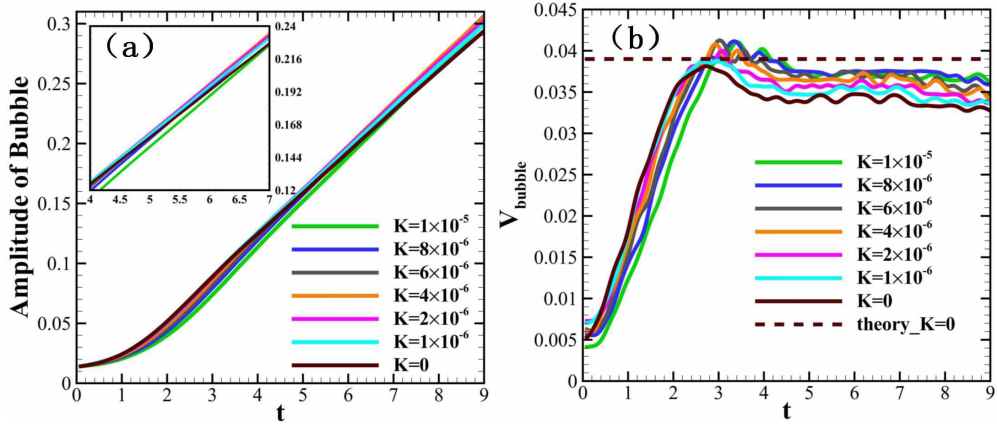}
\caption{Evolution of (a) bubble amplitudes, and (b) bubble velocity over time for different interfacial tension coefficients.} \label{Fig3}
\end{figure}

Figure \ref{Fig3} show the growth of bubble amplitude and bubble velocity over time for different interfacial tension coefficients. It can be seen that the effect of interfacial tension on the development of RTI is stage-specific. At the stage of bubble acceleration, the bubble amplitude and bubble velocity decrease with increasing interfacial tension coefficient, and the interfacial tension exhibits a suppressive effect on RTI development. At the asymptotic velocity stage, the bubble amplitude and bubble velocity tend to first increase and then decrease with increasing interfacial tension coefficient, and the interfacial tension exhibits a first facilitating and then suppressive effect on RTI development.
The brown solid line in Fig. \ref{Fig3}(b) shows the growth of the bubble velocity when the interfacial tension coefficient is $0$.
The brown dashed line shows the corresponding theoretical value of the asymptotic bubble velocity calculated by 
\begin{equation}
{{{u}}_{b}}=\sqrt{\frac{2Atg}{3k\left( 1+At \right)}+\frac{4}{9}{{k}^{2}}{{v}_{h}}^{2}-\frac{\xi K k}{9{{\rho }_{h}}}}-\frac{2}{3}k{{v}_{h}}\text{.}  \label{Eq.22}
\end{equation}%
which is theoretical equation for calculating the asymptotic velocity of the bubble\cite{sohn2009effects,xin2013bubble},The results obtained from the numerical simulation and Eq. \eqref{Eq.22} are in good agreement, demonstrating that the numerical simulation results are correct. The asymptotic bubble velocity obtained from the numerical simulation is lower than the theoretical value. This is mainly due to the fact that the Atwood number ${\rm At}$ at the top of the bubble decreases gradually during the evolution of the RTI, whereas the theoretical value is calculated from ${\rm At}$ at the initial moment. The decrease of ${\rm At}$ leads to a lower asymptotic velocity.

\begin{figure}[tbp]
\center{\epsfig{file=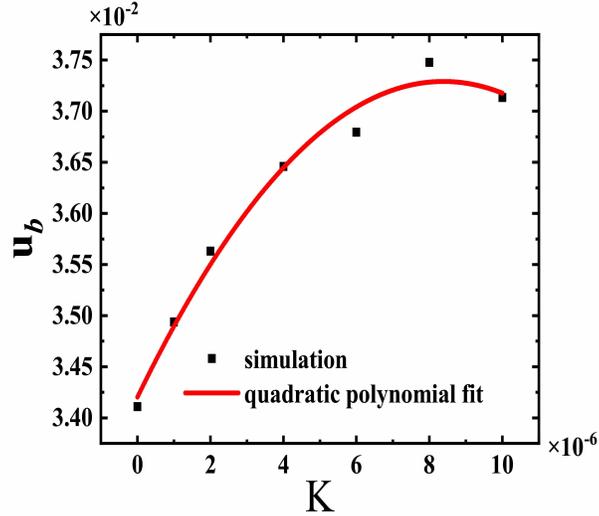,
width=0.5\textwidth,trim=10 10 10 10,clip }}
\caption{Varation of asymptotic bubble velocity with interfacial tension coefficients.} \label{Fig4}
\end{figure}

\begin{figure}[tbp]
\center{\epsfig{file=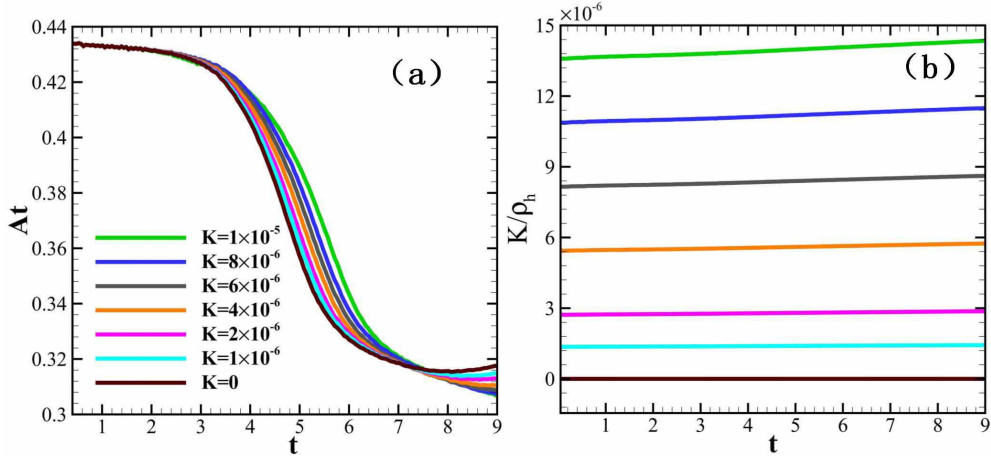,
width=0.8\textwidth,trim=0 0 0 0,clip }}
\caption{Evolution of (a) ${\rm At}$, and (b) ${K }/{{{\rho }_{h}}}\;$ at the top of the bubble over time for different interfacial tension coefficients.} \label{Fig5}
\end{figure}

Figure \ref{Fig4} shows the curve of the asymptotic bubble velocity versus the interfacial tension coefficient. It can be seen that the variation of the asymptotic bubble velocity with the interfacial tension coefficient shows an approximately quadratic variation pattern. Equation (\ref{Eq.22}) can help us comprehend what is going on here.
 It can be seen that the asymptotic bubble velocity is related to the Atwood number ${\rm At}$, the gravitational acceleration $g$, the disturbance wave number $k$, the heavy hydrodynamic viscosity coefficient ${{v}_{h}}$, the heavy fluid density ${{\rho }_{h}}$, and the interfacial tension coefficient $K$. Here $\xi $ is the coefficient in $\sigma {=}\xi K$, where $\sigma$ is the interfacial tension per unit length. The coefficient of heavy hydrodynamic viscosity and the number of perturbation waves are constant in this section, so the variation of the asymptotic bubble velocity is mainly caused by the variation of ${\rm At}$ at the top of the bubble and the variation of ${K }/{{{\rho }_{h}}}\;$. The asymptotic velocity of the bubble increases with increasing ${\rm At}$ and decreases with increasing ${K }/{{{\rho }_{h}}}\;$.
 Figure \ref{Fig5}(a) shows the evolution of ${\rm At}$ at the top of the bubble over time. It can be seen that the ${\rm At}$ increases as the interfacial tension increases. This leads to the increase of the asymptotic bubble velocity with increasing interfacial tension coefficient.
 Figure \ref{Fig5}(b) shows the evolution of ${K }/{{{\rho }_{h}}}\;$ over time. It can be seen that ${K }/{{{\rho }_{h}}}\;$ increases as the interfacial tension increases. This leads to the decrease of the asymptotic bubble velocity with increasing interfacial tension coefficient.
 The two competing factors cause the asymptotic bubble velocity to first increase and then decrease as the interfacial tension factor increases, as shown in  Fig. \ref{Fig4}.

\begin{figure}[tbp]
\center{\epsfig{file=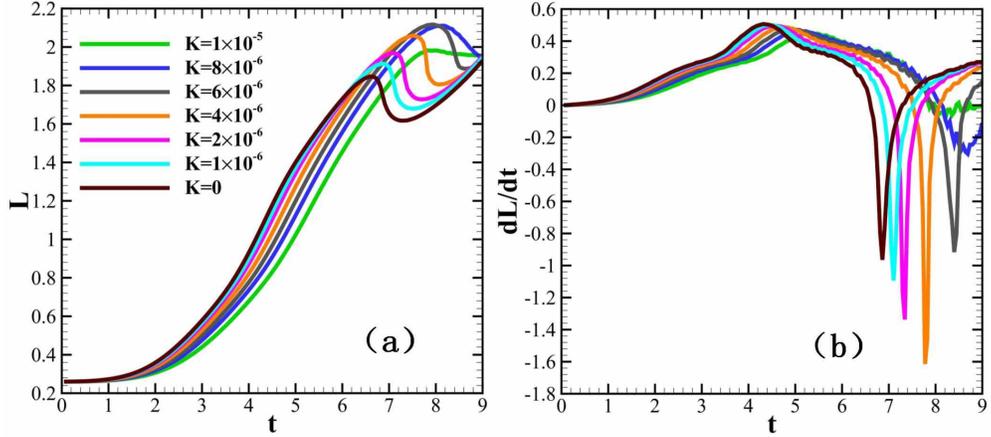,
width=0.8\textwidth,trim=0 0 0 0,clip }}
\caption{Evolution of (a) $L$, and (b) ${dL}/{dt}\;$ over time for different interfacial tension coefficients.} \label{Fig6}
\end{figure}
\begin{figure}[tbp]
\center{\epsfig{file=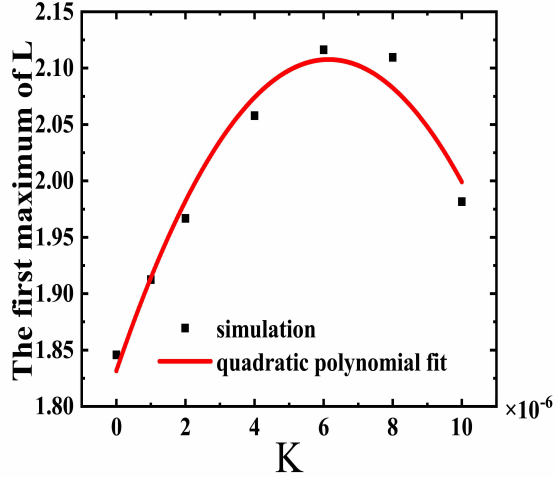,
width=0.5\textwidth }}
\caption{First maximum value of interface length for different interfacial tension coefficients.} \label{Fig7}
\end{figure}

Figure \ref{Fig6}(a) shows the evolution of the interfacial length over time for different interfacial tension coefficients. Figure \ref{Fig7} shows the first maximum value of the interfacial length for different interfacial tension coefficients. It can be seen that before the interfacial length reaches its first maximum, it decreases with the increasing interfacial tension coefficient. The first maximum of the interfacial length tends first to increase and then decrease with the increase of the interfacial tension coefficient. These are due to the different effects of interfacial tension on the growth rate of interfacial length at different stages of RTI development.

Figure \ref{Fig6}(b) shows the evolution of interfacial length change rate over time for different interfacial tension coefficients. It can be seen that the effect of the interfacial tension on the change rate of the interfacial length is stage specific. Before ${dL}/{dt}\;$ reaches its first maximum, the interfacial length change rate decreases with the increase of the interfacial tension coefficient. At this stage, the mushroom structure at the spikes has not yet formed, as shown in Figs. \ref{Fig1}$(a)-1(d)$. The ${dL}/{dt}\;$ is mainly influenced by the movement of spikes into the light fluid and it increase with time. At this stage, ${dL}/{dt}\;$ decreases with increasing interfacial tension coefficient due to the suppressive effect of interfacial tension on the evolution of RTI.
After it reaches its first maximum, the mushroom structure forms. The ${dL}/{dt}\;$ is mainly influenced by the curling of the spikes into the lighter fluid. It decreases with time at this stage. As shown in Fig. \ref{Fig2}, as the interfacial tension increases, the mushroom structure formed at the spike becomes more full and more aggregated. The higher the interfacial tension is, the more heavy fluids are converged at the head of the spike. This facilitates the curling of the spikes into the lighter fluid. This induces the ${dL}/{dt}\;$ increase with increasing interfacial tension coefficient. When the mushroom becomes small, as shown inin Figs. \ref{Fig2}$(g)-2(i)$ , the ${dL}/{dt}\;$ is mainly influenced by the falling of the spikes into the light fluid again. So the ${dL}/{dt}\;$ increases with time and decreases with increasing interfacial tension coefficient again.

\begin{figure}[tbp]
\center{\epsfig{file=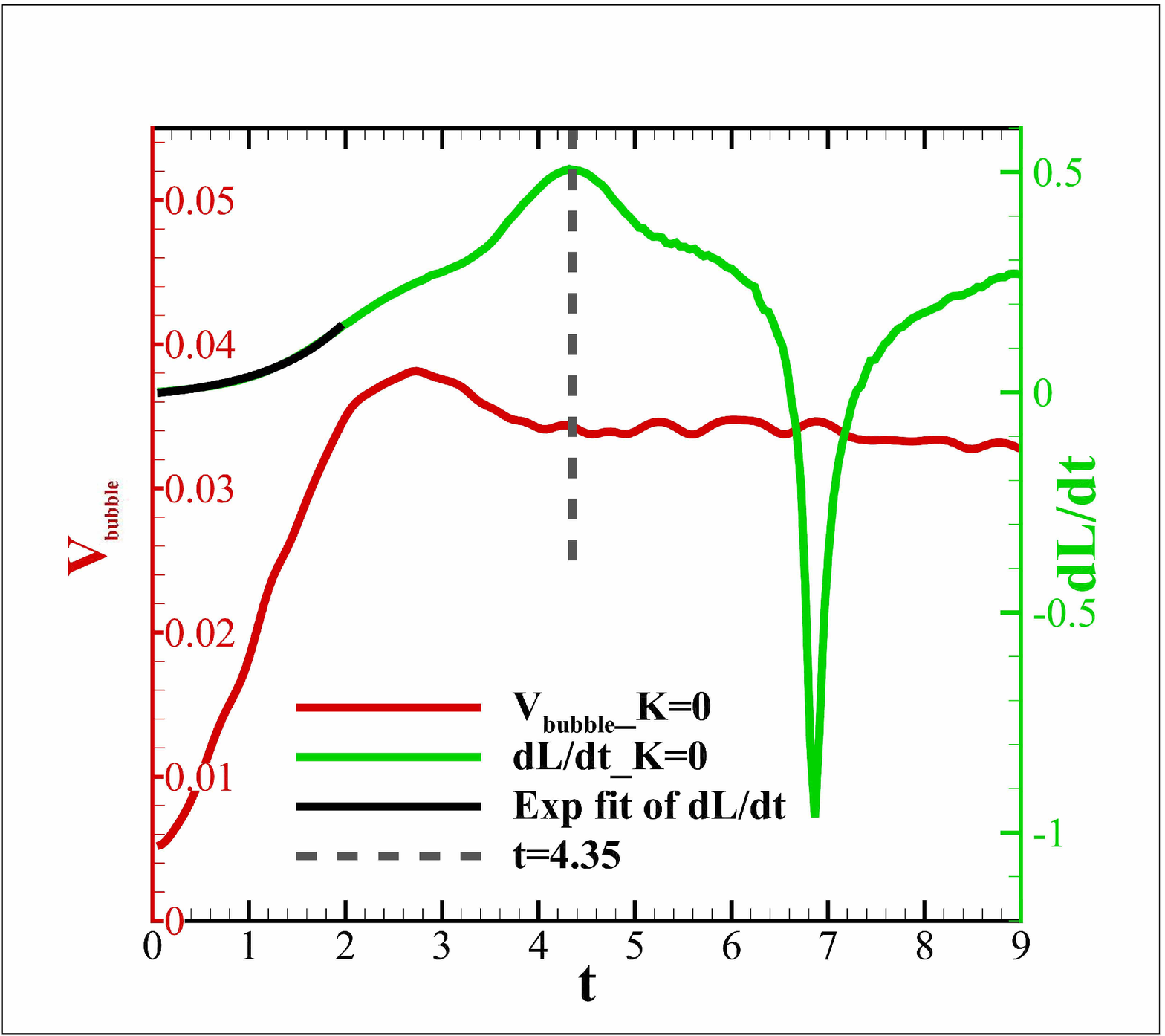,
width=0.5\textwidth,trim=10 10 10 10,clip }}
\caption{Interface length change rate and bubble velocity profiles.} \label{Fig8}
\end{figure}

Figure \ref{Fig8} shows the change rate of the interface length and the bubble velocity profiles. Although the interface length and bubble amplitude are descriptions of RTI development from two different perspectives, there is a correlation between the two. We find that the change rate of the interface length increases exponentially with time in the bubble acceleration stage and the first maximum point of the interface length change rate can be used as a criterion for the development of RTI to enter the asymptotic velocity stage.

\begin{figure}[tbp]
\center{\epsfig{file=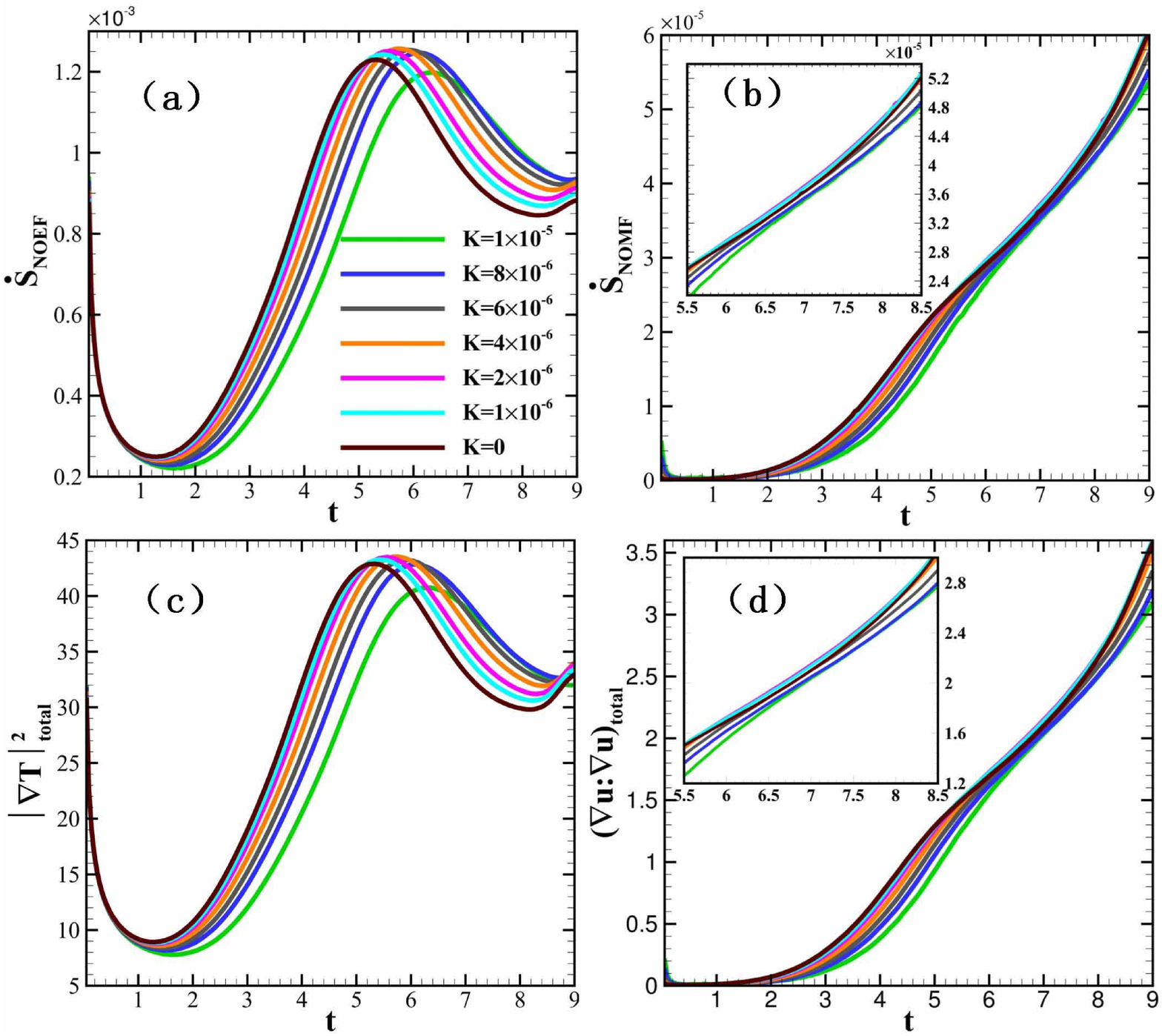,
width=0.8\textwidth,trim=0 0 0 0,clip }}
\caption{Evolution of (a) ${{\dot{S}}_{\text{NOEF}}}$, (b) ${{\dot{S}}_{\text{NOMF}}}$, (c) $\left| \nabla T \right|_{\text{total}}^{2}$, and (d) ${{\left( \nabla \mathbf{u}:\nabla \mathbf{u} \right)}_{\text{total}}}$ over time for different interfacial tension coefficients.} \label{Fig9}
\end{figure}

\begin{figure}[tbp]
\center{\epsfig{file=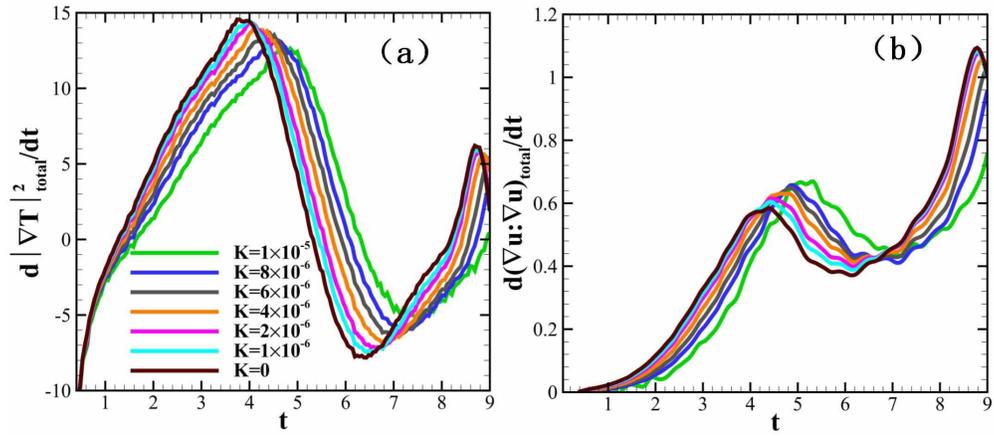,
width=0.8\textwidth,trim=0 0 0 0,clip }}
\caption{Evolution of (a) $d \left| \nabla T \right|_{\text{total}}^{2}/dt$, and (b) $d{{\left( \nabla \mathbf{u}:\nabla \mathbf{u} \right)}_{\text{total}}}/dt$ over time for different interfacial tension coefficients.} \label{Fig10}
\end{figure}
Figures \ref{Fig9}(a) and \ref{Fig9}(b) show the evolution of ${{\dot{S}}_{\text{NOEF}}}$ and ${{\dot{S}}_{\text{NOMF}}}$ respectively. It can be seen that the interfacial tension exhibits different effects on ${{\dot{S}}_{\text{NOEF}}}$ and ${{\dot{S}}_{\text{NOMF}}}$ at different stages. This can be understood with the help of Figs. \ref{Fig9}(c) and \ref{Fig9}(d). As shown in Eqs. (\ref{Eq.16}) and (\ref{Eq.17}), ${{\dot{S}}_{\text{NOEF}}}$ and ${{\dot{S}}_{\text{NOMF}}}$ can be written as functions of temperature and velocity gradients, respectively, in the continuous limit. So the evolution of the flow field total $\left| \nabla T \right|_{\text{total}}^{2}$ and ${{\left( \nabla \mathbf{u}:\nabla \mathbf{u} \right)}_{\text{total}}}$ can, to a certain extent, represent the evolution of ${{\dot{S}}_{\text{NOEF}}}$ and ${{\dot{S}}_{\text{NOMF}}}$, respectively. As shown in Figs. \ref{Fig9}(a) and \ref{Fig9}(c), the evolution of ${{\dot{S}}_{\text{NOEF}}}$ and that of $\left| \nabla T \right|_{\text{total}}^{2}$ are almost the same. Similarly, it can be seen from Figs. \ref{Fig9}(b) and \ref{Fig9}(d) that the evolution of ${{\dot{S}}_{\text{NOMF}}}$ and that of ${{\left( \nabla \mathbf{u}:\nabla \mathbf{u} \right)}_{\text{total}}}$ are almost the same.

At the initial time, the temperature of the heavy fluid is $1.5$ and the temperature of the light fluid is $3.0$. So $\left| \nabla T \right|_{\text{total}}^{2}$ mainly develops from the interface of heavy and light fluids and it is strongly influenced by the change of the interface length.
 Figure \ref{Fig10}(a) shows the evolution of $d \left| \nabla T \right|_{\text{total}}^{2}/dt$ over time for different interfacial tension coefficients. It can be seen that the evolution of $d \left| \nabla T \right|_{\text{total}}^{2}/dt$ is very similar to that of ${dL}/{dt}\;$.
 On the one hand, the increase of the interface length can increase $\left| \nabla T \right|_{\text{total}}^{2}$. The larger ${dL}/{dt}\;$ is, the larger $d \left| \nabla T \right|_{\text{total}}^{2}/dt$ is. On the other hand, heat conduction can lead to a decrease in $\left| \nabla T \right|_{\text{total}}^{2}$. There is a competition between the increase of interface length and the heat conduction.
At an early time, the growth of RTI is very slow and the increase of the interface length is very small. The $\left| \nabla T \right|_{\text{total}}^{2}$ is mainly influenced by the heat conduction at the interface. So $d \left| \nabla T \right|_{\text{total}}^{2}/dt$ is smaller than $0$ and $\left| \nabla T \right|_{\text{total}}^{2}$ decreases with time at this stage. After a short period, the interface instability begins to increase rapidly. So the interface length gets a rapid growth and this induce the increase of $\left| \nabla T \right|_{\text{total}}^{2}$. With the development of RTI, the interface length is much longer, promoting the heat conduction between the heavy fluid and the light one. So $d \left| \nabla T \right|_{\text{total}}^{2}/dt$ can be negative even though the interface length increases just like at $t=6$ as shown in Figs. \ref{Fig6}(b) and \ref{Fig10}(a).

As mentioned before, ${dL}/{dt}\;$ decreases with increasing interfacial tension coefficients at first, then increases with increasing interfacial tension coefficients and decreases with increasing interfacial tension coefficients again. This induces $d \left| \nabla T \right|_{\text{total}}^{2}/dt$ to change with the increasing interfacial tension coefficients in the same way. Consequentially, $\left| \nabla T \right|_{\text{total}}^{2}$ and ${{\dot{S}}_{\text{NOEF}}}$ decrease with increasing interfacial tension coefficients before they reach their maximum and first increase and then decrease with increasing interfacial tension coefficients after that.

\begin{figure}[tbp]
\center{\epsfig{file=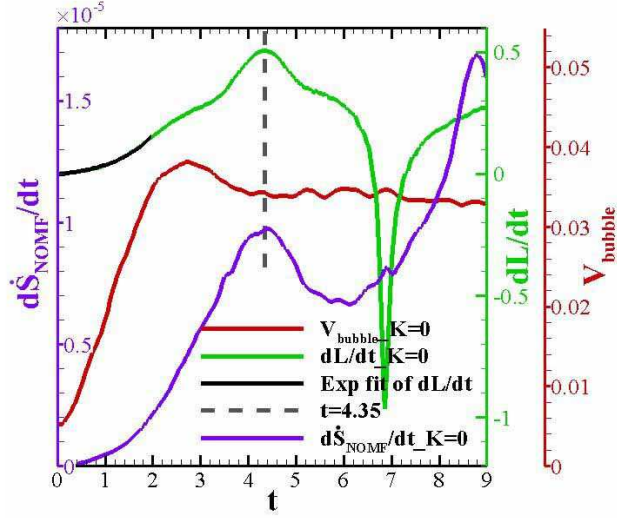,
width=0.5\textwidth,trim=10 10 10 10,clip }}
\caption{Curves of $d {{\dot{S}}_{\text{NOMF}}}/dt$, ${{\mathbf{V}}_{\text{bubble}}}$ and ${dL}/{dt}\;$ over time when the interfacial tension coefficient is zero.} \label{Fig11}
\end{figure}

Similarly, ${{\left( \nabla \mathbf{u}:\nabla \mathbf{u} \right)}_{\text{total}}}$ also mainly develops from the interface of heavy and light fluids. So it is also strongly influenced by the change of the interface length. As shown in Fig. \ref{Fig10}(b), $d{{\left( \nabla \mathbf{u}:\nabla \mathbf{u} \right)}_{\text{total}}}/dt$ is also strongly influenced by ${dL}/{dt}$. The evolution of ${dL}/{dt}$ induce the $d{{\left( \nabla \mathbf{u}:\nabla \mathbf{u} \right)}_{\text{total}}} /dt$ to evolve in the way shown in Fig. \ref{Fig9}(d), which further leads to the evolution of ${{\left( \nabla \mathbf{u}:\nabla \mathbf{u} \right)}_{\text{total}}}$. This is how interfacial tension influences the ${{\dot{S}}_{\text{NOMF}}}$.
By studying and calculating the entropy production rate, it is interesting to find that the first maximum value of $d {{\dot{S}}_{\text{NOMF}}} /dt$ can be used as another criterion for the development of RTI to enter the asymptotic velocity stage, as shown in Fig. \ref{Fig11}.

\subsection{Effect of viscosity on RTI}

\begin{figure}[tbp]
\center{\epsfig{file=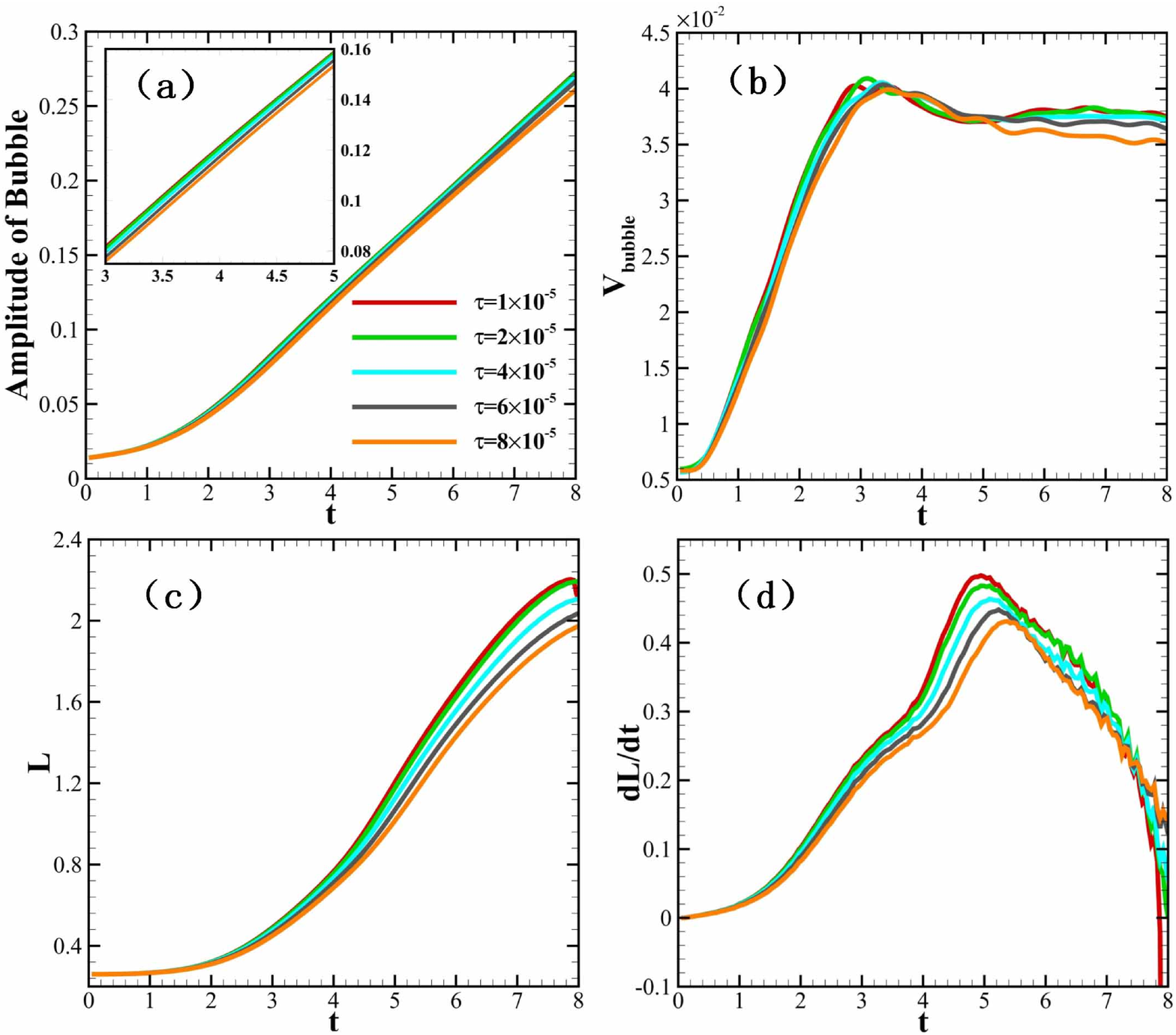,
width=0.8\textwidth,trim=0 0 0 0,clip }}
\caption{Growth of (a) bubble amplitude, (b) bubble velocity, (c) interface length, and (d) interface length change rate over time for different viscosity coefficients.} \label{Fig12}
\end{figure}

In this section, the gravitational acceleration is set to be $g=0.2$ and the interfacial tension coefficient is set to be $K =8\times \text{1}{{\text{0}}^{-6}}$ to simulate the evolution of RTI for different viscosity coefficients to investigate the effect of viscosity on RTI evolution. Various viscosity coefficients are obtained by changing the relaxation time $\tau$ since $\mu =\tau \rho T$. In order to keep the heat conductivity $\kappa ={\mu {{c}_{p}}}/{\Pr}\;$ constant while varying the relaxation time, we simultaneously change the Prandtl number so that the ratio of the viscosity coefficient to the Prandtl number is constant. The bubble amplitude and bubble velocity curves for different viscosity coefficients are shown in Figs. \ref{Fig12}(a) and \ref{Fig12}(b). It can be seen that as the viscosity increases, the bubble amplitude and bubble velocity decrease. Figures \ref{Fig12}(c) and \ref{Fig12}(d) show the evolution of interface length and interface length change rate over time for different viscosity coefficients respectively. With the increase of viscosity coefficient, the interface length and its change rate decrease. Viscosity shows an inhibiting influence on the growth of RTI.

\begin{figure}[tbp]
\center{\epsfig{file=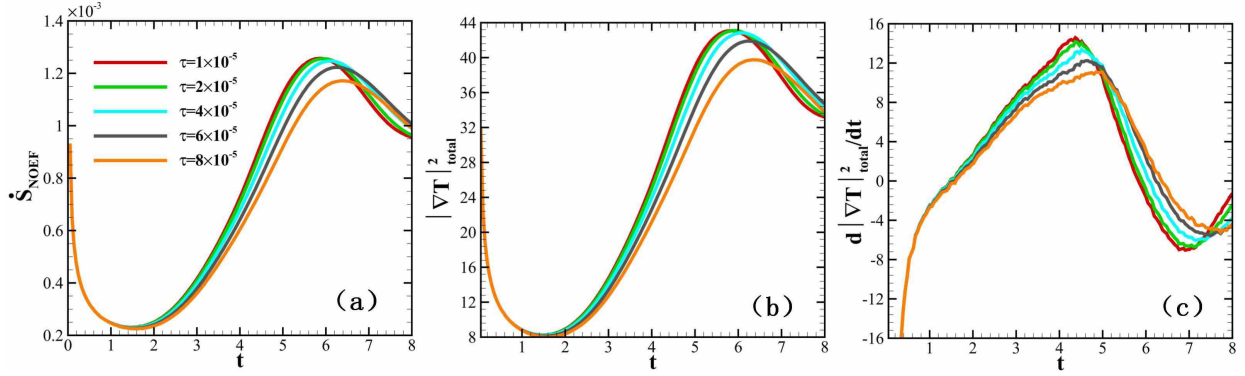,
width=1.0\textwidth,trim=0 0 0 0,clip }}
\caption{Evolution of (a) ${{\dot{S}}_{\text{NOEF}}}$, (b) $\left| \nabla T \right|_{\text{total}}^{2}$, and (c) $d \left| \nabla T \right|_{\text{total}}^{2}/dt$ over time under different viscosity coefficients.} \label{Fig13}
\end{figure}

Figure \ref{Fig13}(a) shows the evolution of the entropy production rates ${{\dot{S}}_{\text{NOEF}}}$ for different viscosity coefficients. The influence of viscosity on ${{\dot{S}}_{\text{NOEF}}}$ can be seen to be stage-specific. Before it reaches its maximum, ${{\dot{S}}_{\text{NOEF}}}$ decreases as viscosity coefficients increase and viscosity exhibits a suppressive effect on ${{\dot{S}}_{\text{NOEF}}}$. After ${{\dot{S}}_{\text{NOEF}}}$ reaches its maximum, ${{\dot{S}}_{\text{NOEF}}}$ shows a trend of first increasing and then decreasing as the viscosity coefficient increases and viscosity exhibits a first facilitating and then suppressive effect on ${{\dot{S}}_{\text{NOEF}}}$. This is mainly because of the competing effects of viscosity on the inhibition of heat convection between light and heavy fluids and the inhibition of RTI development by viscosity. As previously mentioned, ${{\dot{S}}_{\text{NOEF}}}$ is closely related to $\left| \nabla T \right|_{\text{total}}^{2}$.
Figure \ref{Fig13}(b) shows the evolution of $\left| \nabla T \right|_{\text{total}}^{2}$ for different viscosity coefficients. Figure \ref{Fig13}(c) shows the evolution of $d \left| \nabla T \right|_{\text{total}}^{2}/dt$ for different viscosity coefficients. It can be seen that before $d \left| \nabla T \right|_{\text{total}}^{2}/dt$ reaches its maximum, it decreases with the increase of the viscosity coefficient. After $d \left| \nabla T \right|_{\text{total}}^{2}/dt$ reaches its maximum, it increases with the increase of the viscosity coefficient. 

The $\left| \nabla T \right|_{\text{total}}^{2}$ mainly develops from the interface of light and heavy fluids.On the one hand, the increase in viscosity inhibits the development of RTI, which hinders ${dL}/{dt}\;$. On the other hand, the increase in viscosity inhibits the relative movement along the interface, which hinders the heat convection between the light and heavy fluids.
The relative motion along the interface of the light and heavy fluids is relatively weak until $d \left| \nabla T \right|_{\text{total}}^{2}/dt$ reaches its maximum value. The inhibition on the development of RTI lead to the decrease of $d \left| \nabla T \right|_{\text{total}}^{2}/dt$ as the viscosity coefficient increases. This leads to the decrease of $\left| \nabla T \right|_{\text{total}}^{2}$ and ${{\dot{S}}_{\text{NOEF}}}$ with increasing viscosity coefficient. After $d \left| \nabla T \right|_{\text{total}}^{2}/dt$ reaches its maximum value, the relative motion along the interface of the light and heavy fluids is strong. The inhibitory effect of the viscosity on the relative motion becomes essential at this time. This leads to the increase of $d \left| \nabla T \right|_{\text{total}}^{2}/dt$ as viscosity coefficient increases. This is the reason $\left| \nabla T \right|_{\text{total}}^{2}$ and ${{\dot{S}}_{\text{NOEF}}}$ decrease with the increase of the viscosity coefficient at the early time before they reach their maximum and first increase and then decrease with the increase of the viscosity coefficient at the later time after they reach their maximum.

\begin{figure}[tbp]
\center{\epsfig{file=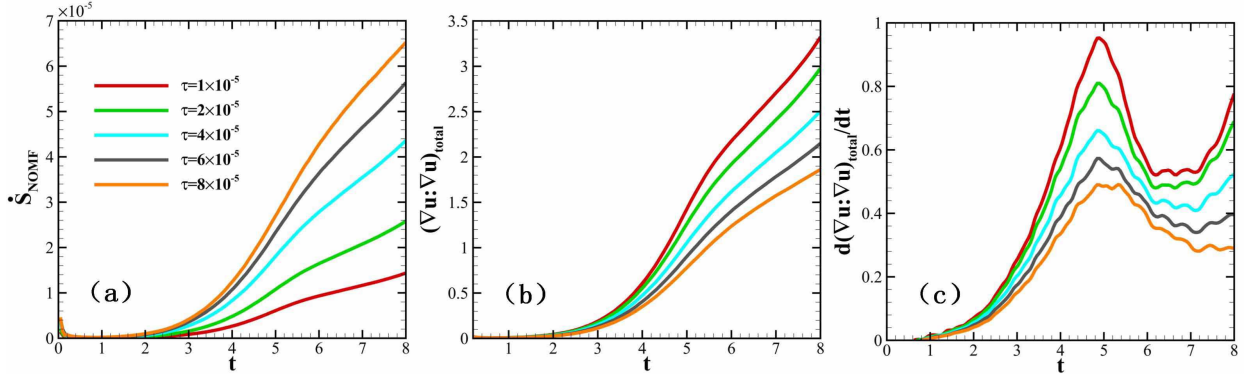,
width=1.0\textwidth,trim=0 0 0 0,clip }}
\caption{Evolution of (a) ${{\dot{S}}_{\text{NOMF}}}$, (b) ${{\left( \nabla \mathbf{u}:\nabla \mathbf{u} \right)}_{\text{total}}}$, and (c) $d{{\left( \nabla \mathbf{u}:\nabla \mathbf{u} \right)}_{\text{total}}}/dt$ over time for different viscosity coefficients.} \label{Fig14}
\end{figure}

Figure \ref{Fig14}(a) depicts the evolution of ${{\dot{S}}_{\text{NOMF}}}$ over time for different viscosity coefficients. It can be observed that ${{\dot{S}}_{\text{NOMF}}}$ grows as viscosity increases, implying that viscosity facilitates ${{\dot{S}}_{\text{NOMF}}}$. This can be understood with the help of Figs. \ref{Fig14}(b) and \ref{Fig14}(c). Figure \ref{Fig14}(b) depicts the evolution of ${{\left( \nabla \mathbf{u}:\nabla \mathbf{u} \right)}_{\text{total}}}$ over time for different viscosity coefficients. Figure \ref{Fig14}(c) shows the development of $d{{\left( \nabla \mathbf{u}:\nabla \mathbf{u} \right)}_{\text{total}}}/dt$ over time for different viscosity coefficients. Both ${{\left( \nabla \mathbf{u}:\nabla \mathbf{u} \right)}_{\text{total}}}$ and $d{{\left( \nabla \mathbf{u}:\nabla \mathbf{u} \right)}_{\text{total}}}/dt$ show a decreasing trend as the viscosity coefficient increases. This is because of the inhibiting influence of viscosity on the growth of RTI. Although the evolution of ${{\dot{S}}_{\text{NOMF}}}$ is close to ${{\left( \nabla \mathbf{u}:\nabla \mathbf{u} \right)}_{\text{total}}}$, it shows a different change trend with the increasing viscosity. As shown in Eq. (\ref{Eq.17}), ${{\dot{S}}_{\text{NOMF}}}$ can be represented as a function of the product of ${{\left( \nabla \mathbf{u}:\nabla \mathbf{u} \right)}_{\text{total}}}$ and the viscosity factor $\mu $. As the viscosity coefficient increases, the product of the two increases. This leads to the increase of ${{\dot{S}}_{\text{NOMF}}}$ as the viscosity coefficient increases.

\subsection{Effect of heat conduction on RTI}

\begin{figure}[tbp]
\center{\epsfig{file=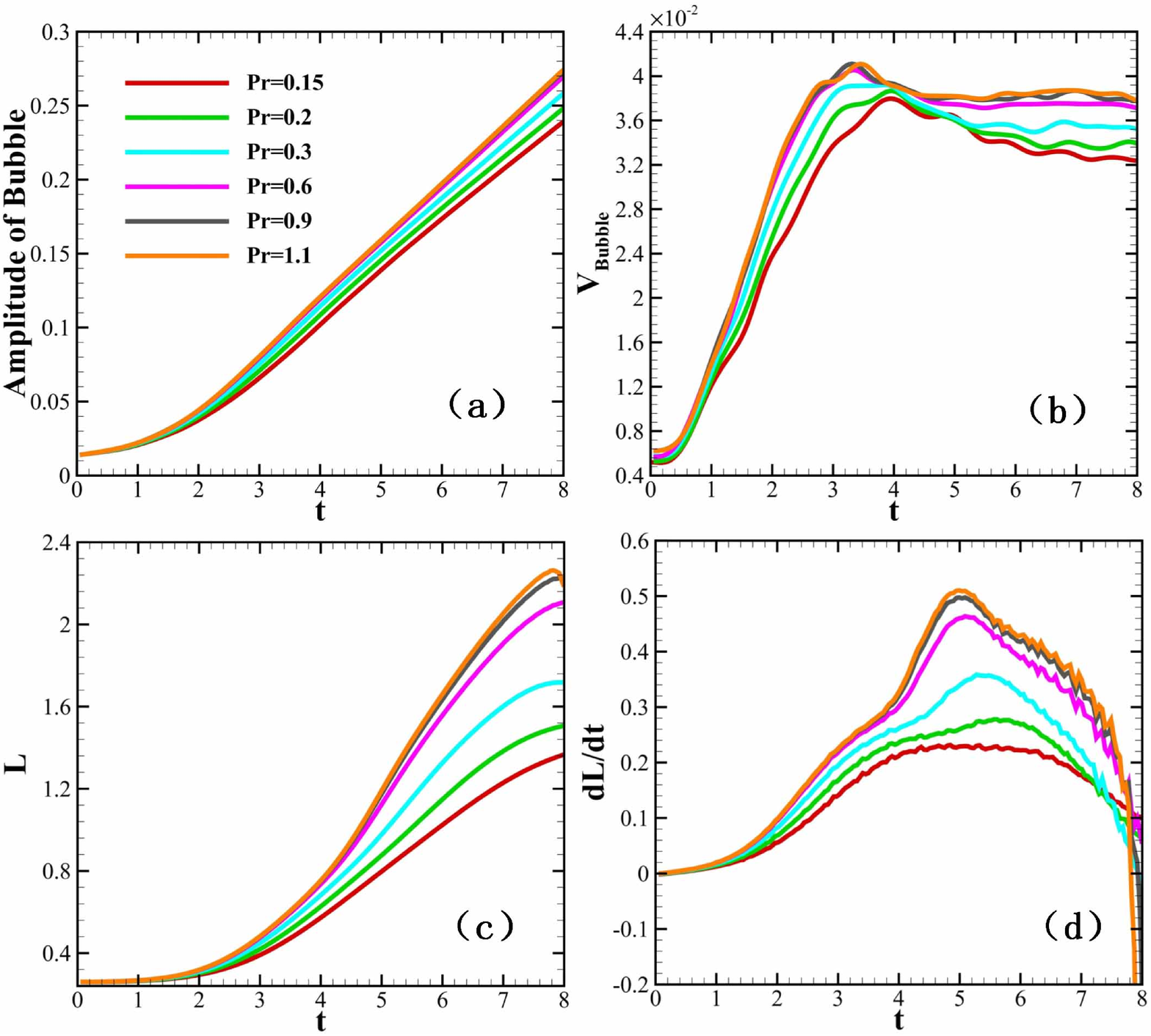,
width=0.8\textwidth,trim=0 0 0 0,clip }}
\caption{Evolution of (a) bubble amplitudes, (b) bubble velocity, (c) interface length, and (d) interface length change rate over time for different heat conductivities.} \label{Fig15}
\end{figure}

In this section, the gravitational acceleration, the relaxation time, and the interfacial tension coefficient are set as $g=0.2$, $\tau =4\times \text{1}{{\text{0}}^{-\text{5}}}$, and $K =8\times \text{1}{{\text{0}}^{-6}}$, respectively. The effect of heat conduction on RTI is investigated by varying the Prandtl number to simulate the evolution of RTI for different heat conductivities. Figures \ref{Fig15}(a) and \ref{Fig15}(b) illustrate the bubble amplitude and bubble velocity curves for various heat conductivities. The bubble amplitude and velocity decrease as the heat conductivity increases. Figures \ref{Fig15}(c) and  \ref{Fig15}(d) depict the evolution of interface length and its change rate over time for different heat conductivities. Both the interface length and its change rate decrease with the increasing conductivities. This leads to the conclusion that heat conduction shows a suppressive influence on the evolution of RTI.

\begin{figure}[tbp]
\center{\epsfig{file=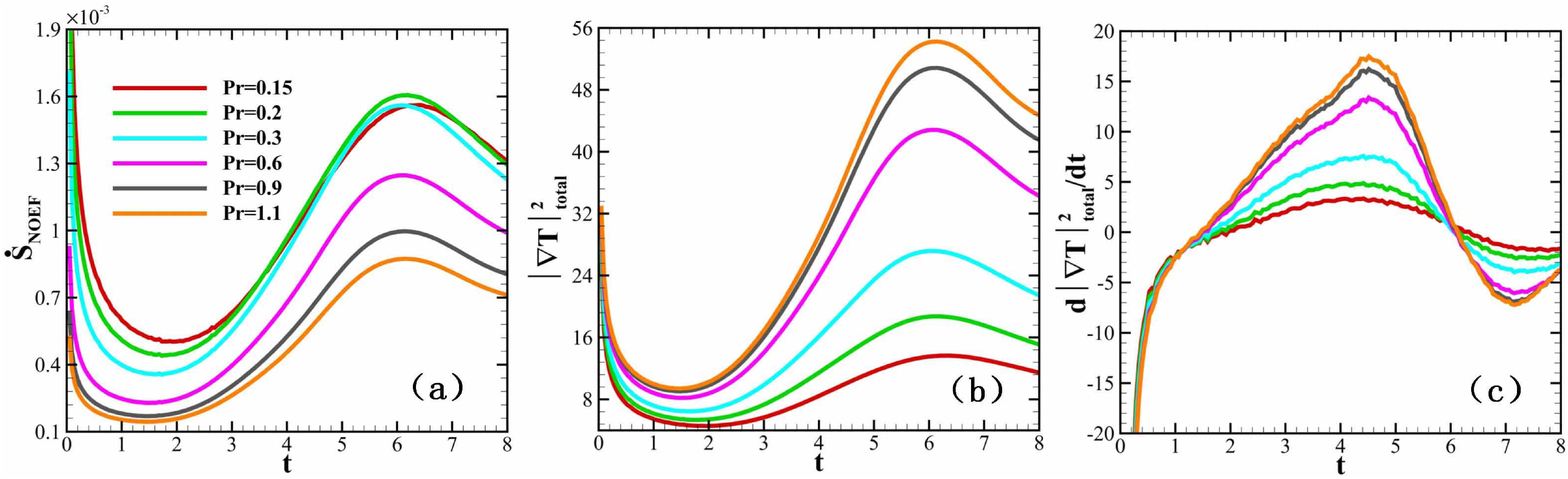,
width=1.0\textwidth,trim=0 0 0 0,clip }}
\caption{Evolution of (a) ${{\dot{S}}_{\text{NOEF}}}$, (b) $\left| \nabla T \right|_{\text{total}}^{2}$, and (c) $d \left| \nabla T \right|_{\text{total}}^{2}/dt$ over time for different heat conductivities.} \label{Fig16}
\end{figure}

Figure \ref{Fig16}(a) depicts the evolution of entropy production rates ${{\dot{S}}_{\text{NOEF}}}$ for various heat conductivities.The ${{\dot{S}}_{\text{NOEF}}}$ increases with increasing heat conductivity at an early stage, but it first increases and then decreases with increasing heat conductivity later. This can be understood with the help of Figs. \ref{Fig16}(b) and \ref{Fig16}(c). Figure \ref{Fig16}(b) depicts the evolution of $\left| \nabla T \right|_{\text{total}}^{2}$ for various heat conductivities. Figure \ref{Fig16}(c) shows the development of $d \left| \nabla T \right|_{\text{total}}^{2}/dt$ for different heat conductivities. As can be seen, $\left| \nabla T \right|_{\text{total}}^{2}$ decreases with increasing heat conductivity.The $d \left| \nabla T \right|_{\text{total}}^{2}/dt$ decreases with increasing conductivity coefficient at an early stage and increase with increasing conductivity coefficient at a later stage.
There is a close correlation between $d \left| \nabla T \right|_{\text{total}}^{2}/dt$ and ${dL}/{dt}\;$. Heat conduction has a suppressive influence on the evolution of RTI. The change rate of the interface length decreases with the increasing conductivity. At an early stage, with the increasing conductivity, ${dL}/{dt}\;$ decreases and the heat conduction is stronger. This leads to $d \left| \nabla T \right|_{\text{total}}^{2}/dt$ decreasing as heat conductivity increases. At a later time, the interface length for lower conductivity is much longer than that for a higher conductivity. The mixing and contact between the light and heavy fluids are sufficient, which promotes the heat transfer between the light and heavy fluids. So $d \left| \nabla T \right|_{\text{total}}^{2}/dt$ for a lower conductivity is smaller than that for a higher conductivity.
At an early time, ${dL}/{dt}$ is small, and $\left| \nabla T \right|_{\text{total}}^{2}$ is mainly influenced by the heat conduction at the interface of the light and heavy fluids. This induces $\left| \nabla T \right|_{\text{total}}^{2}$ to decrease with time at first. With the increasing heat conduction coefficient, the heat conduction is stronger. So $\left| \nabla T \right|_{\text{total}}^{2}$ decrease with the increasing heat conduction coefficient. With the evolution of RTI going, ${dL}/{dt}$ is larger and $L$ is much longer. This induces $\left| \nabla T \right|_{\text{total}}^{2}$ to increase with time. As already stated, the increasing conductivity leads to the decrease of ${dL}/{dt}$ and stronger heat conduction. So the growing rate of $\left| \nabla T \right|_{\text{total}}^{2}$ for a higher heat conductivity is lower. As time goes on, the interface length becomes much longer and the mixing and contact between the light and heavy fluid are more sufficient. So $\left| \nabla T \right|_{\text{total}}^{2}$ decreases with time. With the decrease of heat conductivity, the interface length is much longer and the mixing and contact between light and heavy fluid are more sufficient. This leads to the higher rate of decrease of $\left| \nabla T \right|_{\text{total}}^{2}$ for a lower heat conductivity.
As shown in Eq. (\ref{Eq.16}), ${{\dot{S}}_{\text{NOEF}}}$ can be represented as a function of the product of $\left| \nabla T \right|_{\text{total}}^{2}$ and the heat conductivity $\kappa$. This induces ${{\dot{S}}_{\text{NOEF}}}$ to increase with increasing heat conductivity at an early stage, and ${{\dot{S}}_{\text{NOEF}}}$ first increases and then decreases with increasing heat conductivity at a later time.

\begin{figure}[tbp]
\center{\epsfig{file=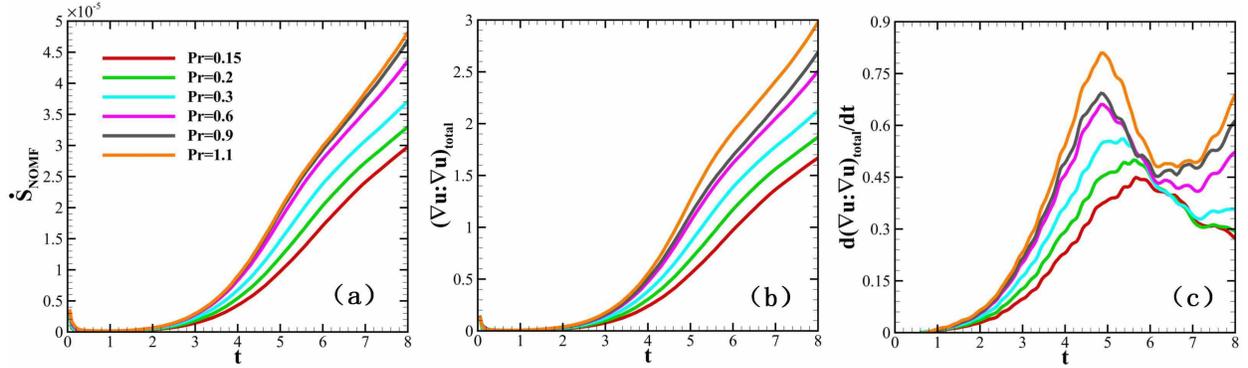,
width=1.0\textwidth,trim=0 0 0 0,clip }}
\caption{Evolution of (a) ${{\dot{S}}_{\text{NOMF}}}$, (b) ${{\left( \nabla \mathbf{u}:\nabla \mathbf{u} \right)}_{\text{total}}}$, and (c) $d{{\left( \nabla \mathbf{u}:\nabla \mathbf{u} \right)}_{\text{total}}}/dt$ over time for different heat conductivities.} \label{Fig17}
\end{figure}

Figure \ref{Fig17} depicts the evolution of entropy production rates ${{\dot{S}}_{\text{NOMF}}}$, ${{\left( \nabla \mathbf{u}:\nabla \mathbf{u} \right)}_{\text{total}}}$, and $d{{\left( \nabla \mathbf{u}:\nabla \mathbf{u} \right)}_{\text{total}}}/dt$ over time for different heat conductivities. As heat conductivity increases, the evolution of RTI is inhibited, the ${dL}/{dt}\;$ and $L\;$ are smaller and the mutual movement between the light and heavy fluids is weaker. This induces $d{{\left( \nabla \mathbf{u}:\nabla \mathbf{u} \right)}_{\text{total}}}/dt$, ${{\left( \nabla \mathbf{u}:\nabla \mathbf{u} \right)}_{\text{total}}}$ and ${{\dot{S}}_{\text{NOMF}}}$ to decrease with increasing heat conductivity.

\section{Conclusions and discussions}

In this paper, a DBM considering the intermolecular interactions was developed and used to investigate the two-dimensional single-mode RTI problem.
The effects of interfacial tension, viscosity, and heat conduction on the evolution of RTI and on the two kinds of entropy production rates related to viscous stress (${{\dot{S}}_{\text{NOMF}}}$) and heat conduction (${{\dot{S}}_{\text{NOEF}}}$) were investigated. For the interfacial tension effect, it was discovered that interfacial tension has a suppressive effect on RTI evolution at the bubble acceleration stage. However, at the asymptotic velocity stage, interfacial tension first facilitates and then inhibits the RTI evolution. The interfacial tension also exhibits different effects on ${{\dot{S}}_{\text{NOMF}}}$ and ${{\dot{S}}_{\text{NOEF}}}$ at different stages. It inhibits ${{\dot{S}}_{\text{NOMF}}}$ and ${{\dot{S}}_{\text{NOEF}}}$ at an early stage and first promotes and then inhibits them at a later time.
For the viscosity and heat conduction effect, it was discovered that both viscosity and heat conduction show a suppressive effect on the evolution of RTI. However, there is a little difference between the effects of viscosity and heat conduction on the two kinds of entropy production rates.
Viscosity promotes ${{\dot{S}}_{\text{NOMF}}}$, but it inhibits ${{\dot{S}}_{\text{NOEF}}}$ at an early stage and first promotes and then inhibits ${{\dot{S}}_{\text{NOEF}}}$ at a later time.
The effects of heat conduction on the two kinds of entropy production rates also show a difference.
It was found that heat conduction inhibits ${{\dot{S}}_{\text{NOMF}}}$ but promotes ${{\dot{S}}_{\text{NOEF}}}$ at the early stage and first promotes and then inhibits ${{\dot{S}}_{\text{NOEF}}}$ later.

In addition, it was found that the morphological interface length can be a helpful complement to a two-dimensional depiction of the RTI development process. The first maximum value point of the interface length change rate can be used as a criterion for RTI evolution to enter the asymptotic velocity stage.
The first maximum value of the change rate of the entropy product rate related to viscous stress can be utilized as an additional criterion for the evolution of RTI to enter the asymptotic velocity stage.

When interfacial tension, viscosity and heat conductivity are present together, viscosity and heat conduction also show an inhibitory effect on RTI development. This is consistent with the results of previous studies when interfacial tension was not considered. However, as the coefficients of viscosity and heat conduction of a fluid are related to the fluid's density and temperature, the introduction of intermolecular forces will lead to the evolution characteristics of density and temperature different from that based on an ideal gas model, thus affecting the viscosity and heat conduction during the evolution of the system. We found that the variation patterns of certain non-equilibrium quantities of the system with viscosity and heat conduction may be influenced by the interfacial tension. The evolution of these non-equilibrium quantities with viscosity and heat conduction maybe different from that without considering the interfacial tension.

The study of RTI problems considering intermolecular forces using the DBM is just in its infancy. In this work we started with a relatively simple problem, a two-dimensional single-mode RTI. The two-dimensional simulation can be regarded as a special case of the three-dimensional problem that is uniform and unchanged in the third dimension Z.
Of course, there will be some differences between three- and two-dimensional calculations. For three-dimensional problems, the evolution of the interface between light and heavy fluids will no longer be described by the interface length, but instead by the interface area. For a simple three-dimensional single-mode RTI problem, there should also be a turning point in the change rate of the interface area that can be used as a criterion for the development of RTI into another stage. Non-equilibrium quantities such as the temperature gradient and velocity gradient of the system, as well as their change rates, will also show different variation patterns at different stages of RTI development. These are similar to the two-dimensional single-mode RTI. However, it is more complicated that even for simple the three-dimensional single-mode situation, there is coupling between perturbations in various directions. This could make the interface's evolution much more complex and should be studied further.

\section*{Acknowledgements}

The authors thank Yanbiao Gan, Beibei Sun, Feng Chen, Huilin Lai, Chuandong Lin, Ge Zhang, Dejia Zhang, Jiahui Song, Yiming San, Hanwei Li, Yingqi Jia, Fanfei Meng and Lifeng Wang for helpful discussions.
This work was supported by the National Natural Science Foundation of China (Grants No.12172061, No.12102397 and No.12101064),  Opening Project of State Key Laboratory of Explosion Science and Technology (Beijing Institute of Technology) under Grant No.KFJJ21-16 M, and Foundation of Laboratory of Computational Physics.

\section*{Appendix A: Discrete Velocity Model}

\begin{figure}[tbp]
\center{\epsfig{file=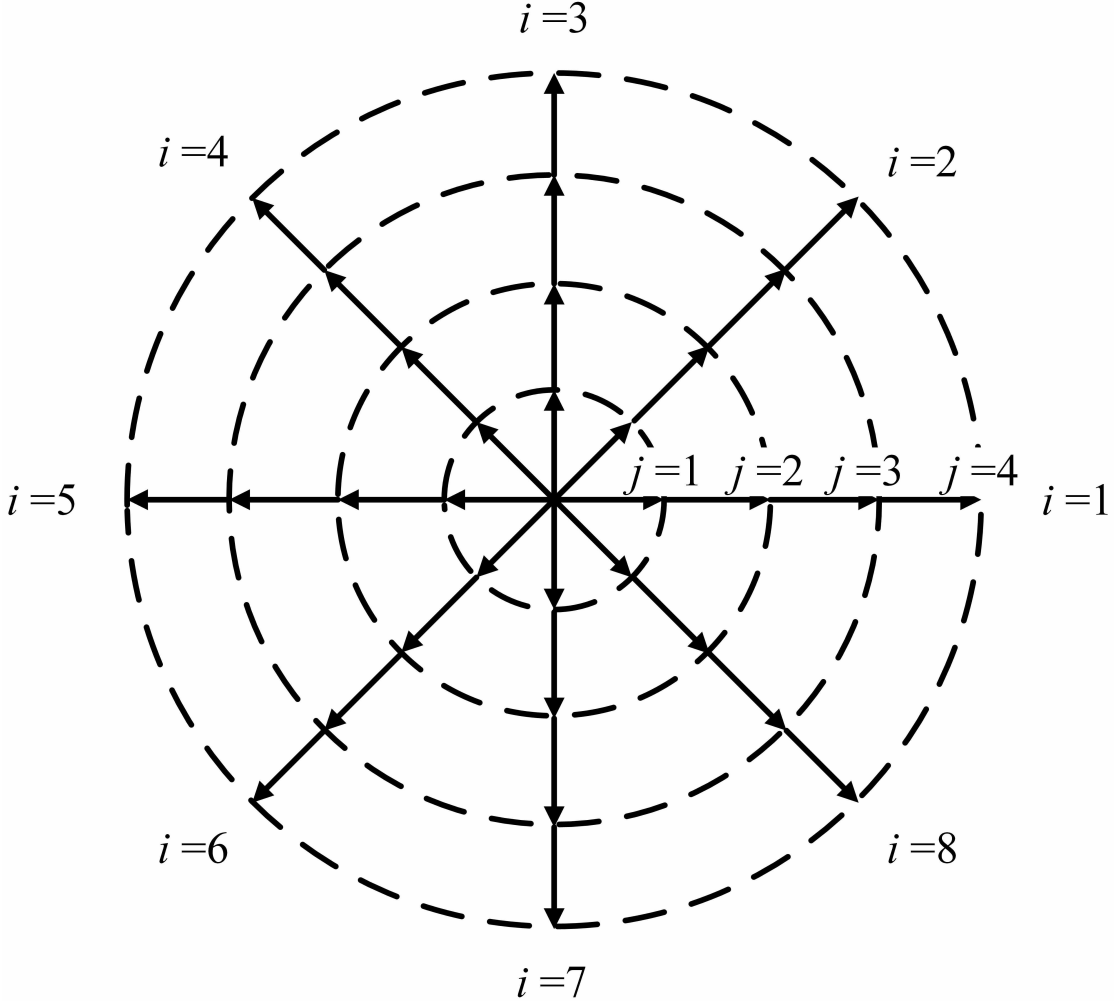,
width=0.9\textwidth,trim=10 10 10 10,clip }}
\caption{Discrete velocity model.} \label{Fig18}
\end{figure}

Figure \ref{Fig18} shows the discrete velocity model used, which contains a zero velocity, as well as four sets of eight velocities in each direction. Each set of discrete velocities has the same velocity value ${{v}_{j}}$; the discrete velocity ${{\mathbf{v}}_{ji}}$ can be written as
\begin{equation}
{{\mathbf{v}}_{ji}}={{v}_{j}}\left[ \cos (\frac{i-1}{4}\pi ),\sin(\frac{i-1}{4}\pi ) \right]\text{,}  \label{Eq.23}
\end{equation}%
where $j=0,1,2,3,4;i=1,2,\cdots ,8$; and $j=0$ is the zero velocity. The equilibrium distribution function corresponding to the discrete velocity can be written as:
\begin{equation}
f_{ji}^{\text{eq}}=\rho {{F}_{j}}\left[ \left( 1-\frac{{{u}^{2}}}{2T}+\frac{{{u}^{4}}}{8{{T}^{2}}} \right)+\frac{{{\mathbf{v}}_{ji}}\cdot \mathbf{u}}{T}\left( 1-\frac{{{u}^{2}}}{2T} \right)+\frac{{{({{\mathbf{v}}_{ji}}\cdot \mathbf{u})}^{2}}}{2{{T}^{2}}}\left( 1-\frac{{{u}^{2}}}{2T} \right) \right.\left. +\frac{{{({{\mathbf{v}}_{ji}}\cdot \mathbf{u})}^{3}}}{6{{T}^{3}}}+\frac{{{({{\mathbf{v}}_{ji}}\cdot \mathbf{u})}^{4}}}{24{{T}^{4}}} \right]\text{,}  \label{Eq.24}
\end{equation}%
where ${{F}_{j}}$ is the weighting factor:
\begin{equation}
{{F}_{1}}=\frac{48{{T}^{4}}-6(v_{2}^{2}+v_{3}^{2}+v_{4}^{2}){{T}^{3}}+(v_{2}^{2}v_{3}^{2}+v_{2}^{2}v_{4}^{2}+v_{3}^{2}v_{4}^{2}){{T}^{2}}-\frac{1}{4}v_{2}^{2}v_{3}^{2}v_{4}^{2}T}{v_{1}^{2}(v_{1}^{2}-v_{2}^{2})(v_{1}^{2}-v_{3}^{2})(v_{1}^{2}-v_{4}^{2})}\text{,}  \label{Eq.25}
\end{equation}%
\begin{equation}
{{F}_{2}}=\frac{48{{T}^{4}}-6(v_{1}^{2}+v_{3}^{2}+v_{4}^{2}){{T}^{3}}+(v_{1}^{2}v_{3}^{2}+v_{1}^{2}v_{4}^{2}+v_{3}^{2}v_{4}^{2}){{T}^{2}}-\frac{1}{4}v_{1}^{2}v_{3}^{2}v_{4}^{2}T}{v_{2}^{2}(v_{2}^{2}-v_{1}^{2})(v_{2}^{2}-v_{3}^{2})(v_{2}^{2}-v_{4}^{2})}\text{,}  \label{Eq.26}
\end{equation}%
\begin{equation}
{{F}_{3}}=\frac{48{{T}^{4}}-6(v_{1}^{2}+v_{2}^{2}+v_{4}^{2}){{T}^{3}}+(v_{1}^{2}v_{2}^{2}+v_{1}^{2}v_{4}^{2}+v_{2}^{2}v_{4}^{2}){{T}^{2}}-\frac{1}{4}v_{1}^{2}v_{2}^{2}v_{4}^{2}T}{v_{3}^{2}(v_{3}^{2}-v_{1}^{2})(v_{3}^{2}-v_{2}^{2})(v_{1}^{2}-v_{4}^{2})}\text{,}  \label{Eq.27}
\end{equation}%
\begin{equation}
{{F}_{4}}=\frac{48{{T}^{4}}-6(v_{1}^{2}+v_{2}^{2}+v_{3}^{2}){{T}^{3}}+(v_{1}^{2}v_{2}^{2}+v_{1}^{2}v_{3}^{2}+v_{2}^{2}v_{3}^{2}){{T}^{2}}-\frac{1}{4}v_{1}^{2}v_{2}^{2}v_{3}^{2}T}{v_{4}^{2}(v_{4}^{2}-v_{1}^{2})(v_{4}^{2}-v_{2}^{2})(v_{4}^{2}-v_{3}^{2})}\text{,}  \label{Eq.28}
\end{equation}%
\begin{equation}
{{F}_{0}}=1-8({{F}_{1}}+{{F}_{2}}+{{F}_{3}}+{{F}_{4}})\text{.}  \label{Eq.29}
\end{equation}%

\section*{Appendix B: Discussion about the small-scale dissipation and numerical resolution}
\begin{figure}[tbp]
\center{\epsfig{file=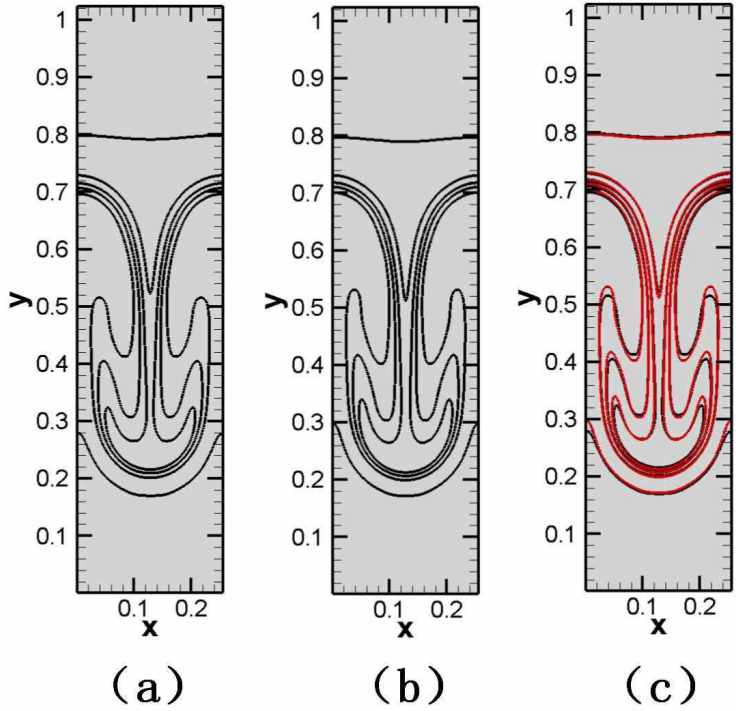,
width=0.6\textwidth,trim=1 1 1 1,clip }}
\caption{Comparison of density at time $t = 6$ for different numerical resolutions: (a) $256\times1024$ meshes, (b) $320\times1280$ meshes, and (c) comparison of the flow fields under the two mesh numbers where black and read lines represent  $256\times1024$ meshes and $320\times1280$ meshes, respectively. Five equally spaced contour lines from ${\rho } = 0.3$ to ${\rho } = 0.7$.} \label{Fig19}
\end{figure}

Figure \ref{Fig19} shows the contour maps of density at $t = 6$ with two different numerical resolutions. In this case, the interfacial tension coefficient is set to be $K =6\times \text{1}{{\text{0}}^{-6}}$, the relaxation time is $\tau =4\times \text{1}{{\text{0}}^{-5}}$, the gravitational acceleration is $g=0.2$, the Prandtl number is $\rm\Pr =0.6$. Figure \ref{Fig19} (a) shows the density of the flow field with a mesh number $256\times1024$, which was used in our study. Figure \ref{Fig19} (b) shows the density of the flow field with a refined mesh of $320\times1280$. Figure \ref{Fig19} (c) is the contrast of the two results, in which the black represents $256\times1024$ meshes and the red represents $320\times1280$ meshes. It can be seen that the flow fields under the two numerical resolutions are substantially identical, only with very small structural variations.
Although the improvement of numerical resolution can further improve the accuracy of small-scale dissipation calculation, the increase of calculation consumption is very huge. This is one of the reasons that $256\times1024$ meshes are adopted in our study. Another important point is that this work mainly focuses on the statistical characteristics of the evolution of various patterns during interface evolution, thus the improvement of the numerical resolution will not affect the validity of our conclusions.

\section*{Appendix C: Cases and parameters}
The cases and the corresponding parameter settings in this paper can be found in Table \ref{table1}.
\begin{table}
\centering
\caption{Cases and parameters.}
\label{table1}
\resizebox{\textwidth}{!}
{
\begin{tabular}
{m{2cm}<{\centering}|m{1cm}<{\centering}|m{2.1cm}<{\centering}|m{1.3cm}<{\centering}|m{1.2cm}<{\centering}|m{2.2cm}<{\centering}|m{2cm}<{\centering}|m{0.9cm}<{\centering}|m{1.7cm}<{\centering}|m{1.3cm}<{\centering}|m{2.2cm}<{\centering}|m{2.8cm}<{\centering}|m{2cm}<{\centering}}

\hline\hline

\multicolumn{1}{c}{}
&\multicolumn{1}{c}{No.}&\multicolumn{1}{c}{$dx$ and $dy$}& \multicolumn{1}{c}{$dt$} & \multicolumn{1}{c}{$\lambda =d$} & \multicolumn{1}{c}{${{y}_{0}}=0.05d$} & \multicolumn{1}{c}{$k=2\pi /\lambda $} & \multicolumn{1}{c}{$g$} & \multicolumn{1}{c}{$\tau$} & \multicolumn{1}{c}{$\rm\Pr$} & \multicolumn{1}{c}{$\mu =\tau \rho T$}  & \multicolumn{1}{c}{$\kappa ={\mu {{c}_{p}}}/{\Pr}\;$} & \multicolumn{1}{c}{$K$}\\

\hline
\multirow{7}{*}{\begin{tabular}[c]{@{}c@{}}Effect of\\interfacial\\tension\end{tabular}} & $1$   & \multirow{18}{*}{$0.001$} & \multirow{18}{*}{$\text{6}\times10\textsuperscript{-6}$} & \multirow{18}{*}{$0.256$} & \multirow{18}{*}{$0.0128$} & \multirow{18}{*}{$24.54$} & \multirow{18}{*}{$0.2$} & \multirow{7}{*}{$\text{4}\times10\textsuperscript{-5}$} & \multirow{7}{*}{$0.6$} & \multirow{7}{*}{$\text{4}{\rho }_{0}{T}_{0}\times10\textsuperscript{-5}$} & \multirow{7}{*}{$\text{6.67}{\rho }_{0}{T}_{0}\times10\textsuperscript{-5}$} & $\text{1}\times10\textsuperscript{-5}$                   \\
                                               & $2$   &                           &                                            &                           &                            &                           &                         &                                           &                        &                                               &                                              & $\text{8}\times10\textsuperscript{-6}$                   \\
                                               & $3$   &                           &                                            &                           &                            &                           &                         &                                           &                        &                                               &                                              & $\text{6}\times10\textsuperscript{-6}$                   \\
                                               & $4$   &                           &                                            &                           &                            &                           &                         &                                           &                        &                                               &                                              & $\text{4}\times10\textsuperscript{-6}$                   \\
                                               & $5$   &                           &                                            &                           &                            &                           &                         &                                           &                        &                                               &                                              & $\text{2}\times10\textsuperscript{-6}$                   \\
                                               & $6$   &                           &                                            &                           &                            &                           &                         &                                           &                        &                                               &                                              & $\text{1}\times10\textsuperscript{-6}$                   \\
                                               & $7$   &                           &                                            &                           &                            &                           &                         &                                           &                        &                                               &                                              & $0$                                        \\
\cline{1-2}\cline{9-13}
\multirow{5}{*}{\begin{tabular}[c]{@{}c@{}}Effect\\of\\viscosity\end{tabular}}           & $1$   &                           &                                            &                           &                            &                           &                         & $\text{1}\times10\textsuperscript{-5}$                  & $0.15$                 & ${\rho }_{0}{T}_{0}\times10\textsuperscript{-5}$                 & \multirow{5}{*}{$\text{6.67}{\rho }_{0}{T}_{0}\times10\textsuperscript{-5}$} & \multirow{5}{*}{ $\text{8}\times10\textsuperscript{-6}$}  \\
                                               & $2$   &                           &                                            &                           &                            &                           &                         & $\text{2}\times10\textsuperscript{-5}$                  & $0.3$                  & $\text{2}{\rho }_{0}{T}_{0}\times10\textsuperscript{-5}$                  &                                              &                                            \\
                                               & $3$   &                           &                                            &                           &                            &                           &                         & $\text{4}\times10\textsuperscript{-5}$                  & $0.6$                  & $\text{4}{\rho }_{0}{T}_{0}\times10\textsuperscript{-5}$                 &                                              &                                            \\
                                               & $4$   &                           &                                            &                           &                            &                           &                         & $\text{6}\times10\textsuperscript{-5}$                  & $0.9$                  & $\text{6}{\rho }_{0}{T}_{0}\times10\textsuperscript{-5}$                  &                                              &                                            \\
                                               & $5$   &                           &                                            &                           &                            &                           &                         & $\text{8}\times10\textsuperscript{-5}$                  & $1.2$                  & $\text{8}{\rho }_{0}{T}_{0}\times10\textsuperscript{-5}$                 &                                              &                                            \\
\cline{1-2}\cline{9-13}
\multirow{6}{*}{\begin{tabular}[c]{@{}c@{}}Effect\\of heat\\conduction\end{tabular}}     & $1$   &                           &                                            &                           &                            &                           &                         & \multirow{6}{*}{ $\text{4}\times10\textsuperscript{-5}$} & $0.15$                 & \multirow{6}{*}{$\text{4}{\rho }_{0}{T}_{0}\times10\textsuperscript{-5}$} & $\text{2.67}{\rho }_{0}{T}_{0}\times10\textsuperscript{-4}$                 & \multirow{6}{*}{$\text{8}\times10\textsuperscript{-6}$}  \\
                                               & $2$   &                           &                                            &                           &                            &                           &                         &                                           & $0.2$                  &                                               & $\text{2}{\rho }_{0}{T}_{0}\times10\textsuperscript{-4}$                                 &                                            \\
                                               & $3$   &                           &                                            &                           &                            &                           &                         &                                           & $0.3$                  &                                               & $\text{1.33}{\rho }_{0}{T}_{0}\times10\textsuperscript{-4}$                                 &                                            \\
                                               & $4$   &                           &                                            &                           &                            &                           &                         &                                           & $0.6$                  &                                               & $\text{6.67}{\rho }_{0}{T}_{0}\times10\textsuperscript{-5}$                 &                                            \\
                                               & $5$   &                           &                                            &                           &                            &                           &                         &                                           & $0.9$                  &                                               & $\text{4.44}{\rho }_{0}{T}_{0}\times10\textsuperscript{-5}$                 &                                            \\
                                               & $6$   &                           &                                            &                           &                            &                           &                         &                                           & $1.1$                  &                                               & $\text{3.64}{\rho }_{0}{T}_{0}\times10\textsuperscript{-5}$                 &                                            \\
\hline\hline
\end{tabular}
}
\end{table}

\section*{References}
%

\end{document}